\documentclass[aps,twocolumn,groupedaddress,nofootinbib,english,nobalancelastpage,floatfix,superscriptaddress]{revtex4-1}

\usepackage[latin1]{inputenc}
\usepackage{sistyle}
\pdfoutput=1
\usepackage{amsmath, amssymb}
\usepackage{graphicx}
\usepackage{color}

\def\GeV{{\rm \ GeV}}

\def\be{\begin{equation}}
\def\ee{\end{equation}}

\newcommand{\Fermi}{\textit{Fermi}}

\def\der{{\rm \ d}}

\begin{document}

\title{Updated cosmic-ray and radio constraints on light dark matter:\\
Implications for the GeV gamma-ray excess at the Galactic center}

\author{Torsten Bringmann}
\email{Torsten.Bringmann@fys.uio.no}
\affiliation{Department of Physics, University of Oslo, Box 1048 NO-0316 Oslo, Norway}

\author{Martin Vollmann}
\email{Martin.Vollmann@desy.de}
\affiliation{II. Institute for Theoretical Physics, University of Hamburg, Luruper Chaussee 149, 22761 Hamburg, Germany}

\author{Christoph Weniger}
\email{C.Weniger@uva.nl}
\affiliation{GRAPPA, University of Amsterdam, Science Park 904, 1090
GL Amsterdam, Netherlands}

\begin{abstract}
The apparent gamma-ray excess in the Galactic center region and inner Galaxy has 
attracted considerable interest, notably because both its spectrum and radial 
distribution are consistent with an interpretation in terms of annihilating dark matter 
particles with a mass of about  $10-40$\,GeV. We confront such an interpretation 
with an updated compilation of various indirect dark matter detection bounds, which 
we adapt to the specific form required by the observed signal. 
We find that cosmic-ray positron data strongly rule out dark matter annihilating to 
light leptons, or 'democratically' to all leptons, as an explanation of the signal. 
Cosmic-ray antiprotons, for which we present independent and significantly 
improved limits with respect to previous estimates, are already in considerable 
tension with DM annihilation to any combination of quark final states; the first set of 
AMS-02 data will thus be able to rule out or confirm the DM hypothesis with high 
confidence. For reasonable assumptions about the magnetic 
field in the Galactic center region,  radio observations independently 
put very severe constraints on a DM interpretation of the excess, in particular for all 
leptonic annihilation channels.
\end{abstract}

\maketitle

\section{Introduction} 

Because of the high expected dark matter (DM) density, the inner part of the Galaxy is 
one of the prime targets for indirect searches for particle DM (for recent reviews, see 
Refs.~\cite{Cirelli:2012tf,Lavalle:2012ef,Bringmann:2012ez}). Indeed,  
indications for DM 
signals from the Galactic center (GC) region have repeatedly appeared in the past -- in 
particular in gamma rays 
\cite{Cesarini:2003nr, Horns:2004bk, Bergstrom:2004cy, Bringmann:2012vr, 
Weniger:2012tx, Su:2012ft, Belikov:2012ty}, 
but also in microwaves \cite{Hooper:2007kb} and the annihilation radiation from 
positrons \cite{Boehm:2003bt}. This part of the Galaxy, however,  is 
also an astrophysically very rich environment. Unless one can identify a distinct 
spectral signature \cite{Bergstrom:1997fj, Bringmann:2011ye}, this generally makes 
disentangling any potential DM signal from astrophysical backgrounds a formidable 
task. In fact, more 
refined analyses and new data have so far always tended to disfavour the DM 
hypotheses previously put forward \cite{Lingenfelter:2009kx, Cumberbatch:2009ji, 
Aharonian:2006wh, Stecker:2007xp, Ackermann:2013uma}.  
The field has matured significantly in 
recent years however, concerning both the understanding of astrophysical  
backgrounds and the statistical analysis of potential signals, not the least because 
of the unprecedented wealth of high-quality data that now are at our disposal. At the 
same time, many theoretical models predict signals just below current exclusion 
limits. This implies both that more signal claims should be expected in the near 
future, and that those should not be dismissed too easily. 

An example for such a sophisticated analysis is the recent discussion of the
GeV gamma-ray excess in Fermi-LAT \cite{LAT}
data of the GC and inner Galaxy
\cite{Daylan:2014rsa}, which reconfirms corresponding earlier findings
\cite{Goodenough:2009gk, Hooper:2010mq, Hooper:2011ti, Abazajian:2012pn, 
Hooper:2013rwa, Huang:2013pda, Macias:2013vya, Abazajian:2014fta} with a high 
level of detail (see also Ref.~\cite{Hooper:2012ft} for a review).
The spectrum of the excess seen at the GC can 
be well described by the annihilation of $10-40$\,GeV DM particles into quarks,
or leptons, which then produce secondary photons during their 
fragmentation and decay. Indeed, the signal normalization is consistent with that 
expected for thermally produced DM. The observed spectrum, however, does not contain 
the type of sharp features that would unambiguously point to a DM origin. It is thus 
arguably even more interesting that the signal is spherically symmetric and that it 
is claimed to extend to at least 1.5\,kpc away from the GC~\cite{Hooper:2013rwa}, where 
astrophysical backgrounds with a similar morphology are expected to be strongly 
suppressed. Last but not least, the excess emission decreases with galactocentric 
distance in a way that is consistent with the wide range of expectations for 
annihilating DM. Not surprisingly, the most recent analysis of this excess 
\cite{Daylan:2014rsa} has already triggered a considerable activity in concrete 
model-building attempts \cite{Hektor:2014kga,Agrawal:2014una,Cerdeno:2014cda,
Ipek:2014gua,Ko:2014gha,Boehm:2014bia,Abdullah:2014lla,Ghosh:2014pwa,
Martin:2014sxa,Berlin:2014pya,Basak:2014sza,Cline:2014dwa}.

If the excess can indeed be attributed to annihilating DM particles, this will 
also
leave traces in other cosmic-ray fluxes. In fact, a  confirmation of the DM 
hypothesis essentially {\it requires} to find corroborating evidence from a different 
type of experiment, and observations using other cosmic-ray messengers or
photons at other wavelengths seem to be a particularly natural choice to look for such a second 
signal (while the translation to expected rates in direct 
detection experiments or at colliders is much more model-dependent).
Given the renewed interest in the GC gamma-ray excess, we thus present here an 
updated collection of constraints deriving from indirect DM searches. In particular, 
we point out that the observed radial distribution of the gamma-ray signal essentially 
fixes the DM distribution in the inner Galaxy, thereby significantly reducing the 
astrophysical uncertainties typically associated with such limits. This allows to 
reliably constrain the DM interpretation of the signal, almost 
independent of the DM profile and much less model-dependent than corresponding 
constraints derived from DM searches at colliders or in direct detection 
experiments \cite{Alves:2014yha,Berlin:2014tja,Izaguirre:2014vva,Kong:2014haa,Han:2014nba}.

This article is organized as follows. We start in Section \ref{sec:signal} by
reviewing the current situation of the GeV excess, with an emphasis on the
possibility that it is induced by annihilating DM. In Section 
\ref{sec:constraints} we present updated bounds on this interpretation 
from other indirect searches for DM, in particular using cosmic-ray antiprotons and 
positrons, as well as radio observations. We provide a more detailed discussion
of systematic uncertainties for the most relevant limits in 
Section \ref{sec:discussion}, and a
summary in terms of the main possible DM annihilation channels in context of
the GC GeV excess in Section \ref{sec:summary}. In Section \ref{sec:concl}, finally, we conclude.

\section{Characteristics of the GeV excess}
\label{sec:signal}

The DM interpretation of the excess emission at the GC is suggestive in a
number of ways:  The extended emission appears to peak relatively sharply at energies
around 1--3 GeV, is rotationally symmetric around the GC, and roughly follows a 
$\propto r^{-2.5}$ emission profile, compatible with the annihilation signal from a 
standard cuspy -- if slightly contracted -- DM distribution~\cite{Hooper:2010mq}.  
Furthermore, the same spectral signature was claimed to extend to much higher 
Galactic latitudes, $|b|\gtrsim 10^\circ$~\cite{Hooper:2013rwa} (see also
Ref.~\cite{Huang:2013pda}).
Indeed, a signal from DM annihilation is in general expected to extend to high
latitudes, and can even be visible at the Galactic poles if the substructure
enhancement of the annihilation signal is significant at cosmological
distances.  
Astrophysical explanations in terms of milli-second pulsars (MSPs),
bremsstrahlung or neutral pion decay close to the GC are not yet excluded, but
would all face serious challenges if -- despite the sizeable systematic
uncertainties \cite{Calore:2014} -- the extension of the GeV excess to
Galactic latitudes of $|b|>\mathcal{O}(10^\circ)$ is confirmed.
It should in fact be stressed that the nominal statistical significance of the excess is
extremely high (\textit{e.g.}~at the level of $\sim 40\sigma$ in the inner Galaxy
analysis of Ref.~\cite{Daylan:2014rsa}), and that by now background modeling
uncertainties are the main limiting factor in characterizing its properties.

First claims that the gamma-ray emission from the GC as seen by Fermi LAT
suggests a DM annihilation signal were put forward in
Ref.~\cite{Goodenough:2009gk}, using a simple power-law model for the spectrum
of Galactic diffuse emission.  The authors found that annihilation into
$\bar{b}b$ final states, a DM mass around 25--30 GeV and an annihilation
cross-section close to the thermal value are compatible with the data.  In
Ref.~\cite{Hooper:2010mq} the same authors found that the dominant part of the
emission comes from the inner $1.25^\circ$ around the GC, with a volume
emissivity that scales like $\sim r^{-2.5}$, and discussed possible astrophysical
interpretations in terms of MSPs (first mentioned in
Ref.~\cite{Abazajian:2010zy}; see also Ref.~\cite{Wang:2005av} for a much
earlier discussion in the context of EGRET measurements) and neutral pion decay
from cosmic-ray interactions.  Ref.~\cite{Hooper:2011ti} presented an updated
analysis, using three years of Pass 7 Fermi LAT data, and arriving at similar
conclusions.  Ref.~\cite{Boyarsky:2010dr} pointed out the importance of a
proper treatment of point sources close to the GC.

In an independent analysis, Ref.~\cite{Abazajian:2012pn} confirmed the
existence of a significant extended emission at the GC, and found
it well compatible with the spectrum of known MSPs.
Alternative scenarios discussed in the literature are bremsstrahlung
of electrons on molecular gas~\cite{YusefZadeh:2012nh} within the inner few
hundred pc, or proton-proton interaction within the inner few pc of the
super-massive black hole (SMBH)~\cite{Linden:2012iv}.  In
Ref.~\cite{Gordon:2013vta}, the authors study some of the systematic uncertainties
related to standard emission models for the diffuse backgrounds at the Galactic
center, and find that -- after marginalizing over point sources and diffuse
background uncertainties -- both DM annihilation and a population of at least
$\sim1000$--$2000$ MSPs are
compatible with the excess emission from the Galactic center.

The latest analyses of the GC excess emission were presented in
Ref.~\cite{Abazajian:2014fta}, discussing in some detail background modeling
systematics related to point sources, molecular gas and generic extended
diffuse emission, and in Ref.~\cite{Daylan:2014rsa}, which updates previous
analyses by using a subset of Fermi LAT data with improved angular resolution
(based on a cut on \texttt{CTBCORE}~\cite{Portillo:2014ena}).  For definiteness, we will mostly base the
discussion in this paper on the results obtained in Ref.~\cite{Daylan:2014rsa}.

\subsection{DM interpretation}
\label{sec:gevDM}

Focussing on the DM interpretation of the GeV excess, let us first have a detailed
look at what the observations would tell us about the annihilating particles.
For definiteness, we will do this for a
number of benchmark scenarios -- based on the results from
Ref.~\cite{Daylan:2014rsa}, but additionally including uncertainties related to the DM
distribution.

The differential intensity of photons from DM annihilation as observed at Earth
can be calculated via
\begin{equation}
    \label{eq:qdef}
    \frac{d\phi}{dEd\Omega} = \frac{1}{4\pi}
    \int_{\rm l.o.s.} \!\!\!ds\; 
    \underbrace{\frac{\langle \sigma v\rangle} {2 m_\chi^2}
    \frac{dN_\gamma}{dE}\rho(r)^2}_{\equiv Q(r,E)}\;,
\end{equation}
where $\langle \sigma v\rangle$ is the average velocity-weighted annihilation
cross-section, $m_\chi$ denotes the DM mass, $dN_\gamma/dE$ the energy spectrum of
prompt photons produced in the annihilation, and $\rho(r)$ is the DM density as
function of the galactocentric radius $r$.  The integral runs over the
line-of-sight parameter $s$, and $r$ is given by $r = \sqrt{(R_\odot - s
\cos\psi)^2 + (s \sin \psi)^2}$, where $R_\odot=8.5\rm \ kpc$ is the distance
between Sun and GC, and $\psi$ the angle to the GC.  Lastly, we defined $Q(r,E)$
as the \emph{differential injection rate} of gamma rays from DM annihilation.

When extrapolating the excess emission observed at the GC to other points in
the Galaxy, the main unknown is the shape of the Galactic DM halo and the
distribution of DM substructures.  The infall of baryons onto the central
regions during galaxy formation can cause adiabatic contraction of the DM
halo~\cite{Gnedin:2004cx}.  The exact strength of this effect is unknown, and
we will here use a generalized Navarro-Frenk-White (NFW)
profile~\cite{Navarro:1995iw}, given by
\begin{equation}
\label{nfw}
    \rho(r) = \rho_\odot 
    \left(\frac{r}{R_\odot}\right)^{-\Gamma}
    \left(\frac{r+r_s}{R_\odot+r_s}\right)^{\Gamma-3}\;,
\end{equation}
where $\Gamma$ is the inner slope of the DM halo, and $\rho_\odot$ the
DM density at the position of the Solar system.  Observational constraints from
microlensing and rotation curves of stars or gas~\cite{Iocco:2011jz} on the
slope of the DM halo in the inner few kpc of the Galaxy remain
relatively weak and allow values up to $\Gamma\sim 1.5$.  Throughout,
we will use $\rho_\odot=0.3\rm\ GeV\ cm^{-3}$ and $r_s=20$\,kpc as reference
values.
Note that we are eventually only interested in the {\it ratio} of various constraints to the 
putative signal,  implying that the local DM density $\rho_\odot$ drops out and thus can 
be fixed to any value simply as a matter of convention.  In that case, the variation of the 
annihilation rate at the Sun's position corresponds to a variation of the annihilation cross-section itself. 

In Ref.~\cite{Daylan:2014rsa}, the authors find slopes
$\Gamma\simeq 1.26\pm 0.05$ (at $3\sigma$ CL) from an analysis of the inner
Galaxy (excluding the inner one degree above and below the Galactic disc), and a
value of $\Gamma\simeq 1.04$--$1.24$ from an analysis of the GC source.  We
will quote our main results using the central value of the inner Galaxy
analysis, $\Gamma=1.26$, and will comment on the impact of shallower profiles
when necessary.  Note that throughout the analysis, we will neglect the effect
of substructure in the DM halo. Due to tidal forces, the associated boost at
the GC is in general expected to be negligible; at kpc distances it can however
lead to $\mathcal{O}(1)$ enhancements of the effective annihilation rate (see
\textit{e.g.}~Ref.~\cite{Pieri:2009je}).  Neglecting these effects renders our
constraints conservative.  \medskip

\begin{figure}[t]
    \begin{center}
        \includegraphics[width=\linewidth]{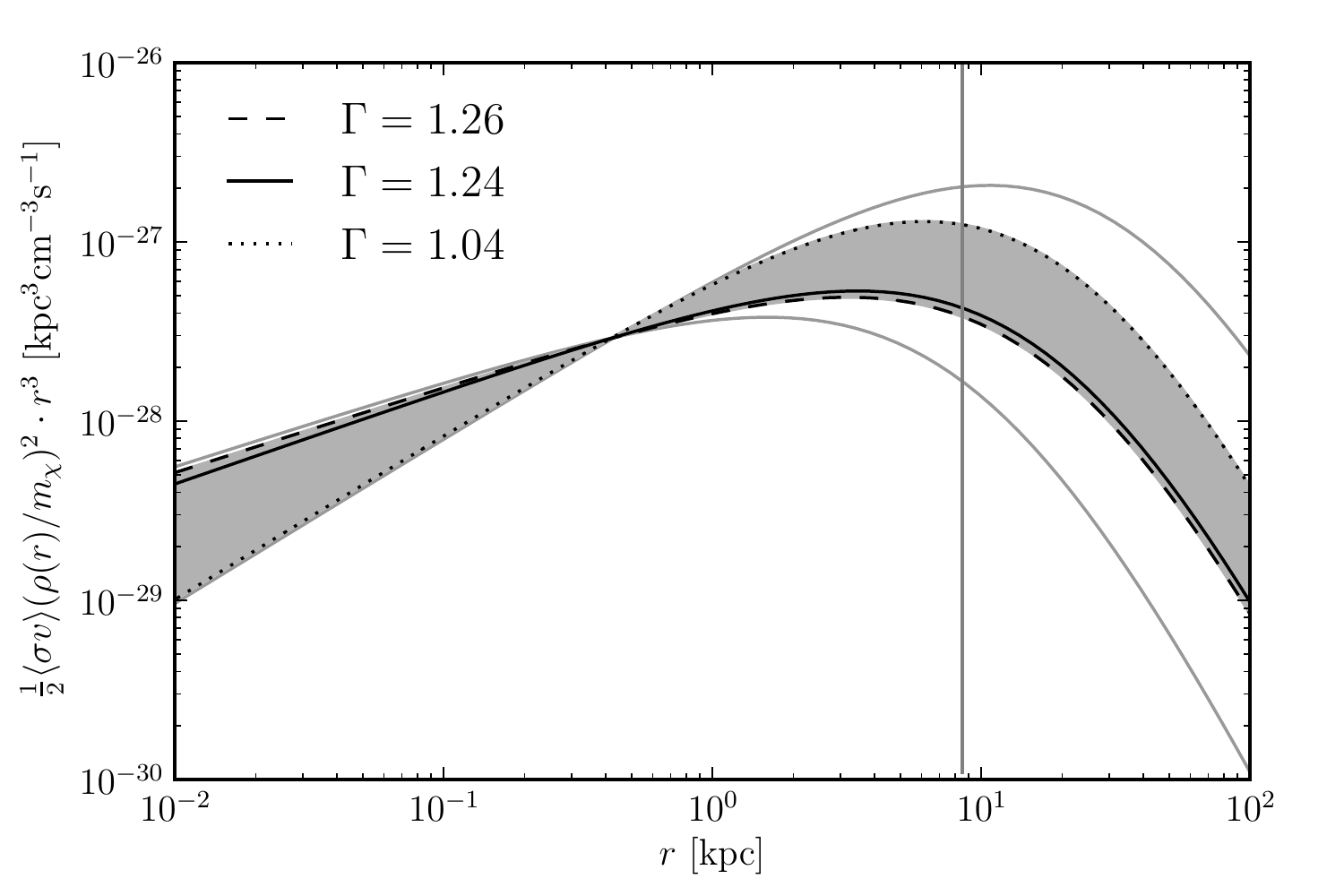}
    \end{center}
     \caption{DM annihilation rate per volume as function of the galactocentric radius $r$,
        for different values of the inner slope $\Gamma$.  The normalization is
        fixed to produce a signal emission in an annulus of
        $1^\circ$--$3^\circ$ around the GC that is compatible with the values
        quoted in Ref.~\cite{Daylan:2014rsa} (see text for details).  At the
        position of the Sun (vertical line) the annihilation
        rate can vary by a factor of 3.3, as indicatd by the black lines.  This
        region is extended as shown by the gray lines when allowing for variations
        in the scale radius of the DM profile, leading to an additional factor
        two uncertainty in both directions.  We multiplied the annihilation
        rate by $r^{3}$ for visual convenience.}
        \label{fig:gamma-dependence}
\end{figure}

We show in Fig.~\ref{fig:gamma-dependence} the radial dependence of the
\textit{DM annihilation rate per volume} for different values of $\Gamma$. The rates are
normalized to yield an identical projected signal flux from the inner
{$1$--$3^\circ$} around the GC, taking $\langle\sigma v\rangle = 1.7\times
10^{-26}\rm \ cm^3 s^{-1}$ into $\bar bb$, $m_\chi=35\rm \ GeV$ and $\Gamma=1.26$ as
benchmark.\footnote{{We checked that normalizing instead in the range
1--2$^\circ$ (1--5$^\circ$) would change the fluxes corresponding to
$\Gamma=1.04$ at most by $+10\%$ ($-15\%$).}}  For slopes $\Gamma$ compatible with the GeV excess, and assuming a
scale radius of $r_s=20\rm\ kpc$, the annihilation rate can vary by about
a factor of $\sim3.3$ (namely, when taking $\Gamma=1.24$ as reference, 2.9 up and 1.1 down). 

Considering possible variations in the
scale radius in the range $r_s=20^{+15}_{-10}\rm\ kpc$, as suggested by DM-only
simulations on the one and dynamical observations on the other hand (see
discussion in Ref.~\cite{Iocco:2011jz}) allows for an additional change of up
to a factor of two in the DM annihilation rate at the position of the Sun (see
Fig.~\ref{fig:gamma-dependence}).
Adopting the proceedure discussed in Ref.~\cite{Cirelli:2010xx} and requiring that a given DM
profile with fixed $\rho_\odot$ and $\Gamma$ can reproduce the SDSS mass constraint $M(r={60\rm\
kpc})=4.7\times10^{11} M_\odot$~\cite{Xue:2008se}, we find that 
$r_s = 24 (34)\rm\ kpc$ for $\Gamma=1.0 (1.26)$ if $\rho_\odot=0.3\rm\ GeV/cm^3$, and
$r_s = 15 (18)\rm\ kpc$ if $\rho_\odot=0.4\rm\ GeV/cm^3$.  Interestingly, these values favour the
upper range of the above uncertainty band.  
Note that uncertainties in $R_\odot$ lead to variations in the annihilation
rate at the Sun's position of at most a few $10\%$, which can be neglected in the present discussion.

Finally, Fig.~\ref{fig:gamma-dependence} demonstrates that steepening the
profile, by increasing $\Gamma$, makes local messengers (like positrons) a
weaker probe of the GeV excess -- while messengers originating from very small
galactocentric distances (like radio signals) will lead to increasingly tighter
constraints -- and {\it vice versa}.  Furthermore, constraints on the GC GeV
excess derived from observations at galactocentric distances of
$\lesssim100\rm\ pc$ (like our radio constraints) are practically independent of
$r_s$.  Combining information from indirect DM searches with different
messengers thus allows to test the DM hypothesis in a way that is even more
independent of the assumed DM distribution than what can be inferred from gamma-ray
observations alone.

\medskip

\begin{figure}
    \begin{center}
        \includegraphics[width=\linewidth]{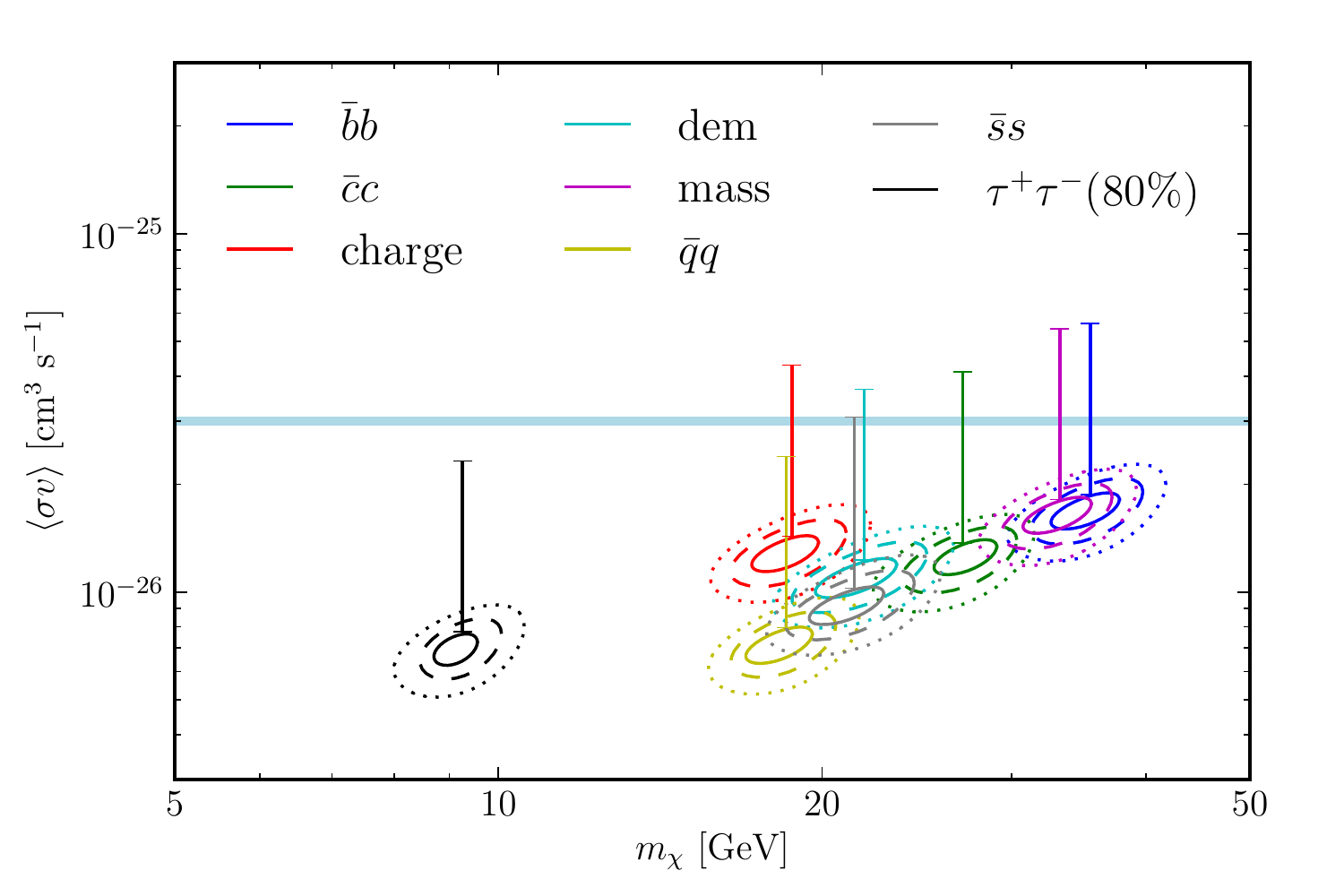}
    \end{center}
    \caption{The \textit{ellipses} show the preferred values of the DM
        annihilation cross-sections and mass from the inner Galaxy analysis of
        Ref.~\cite{Daylan:2014rsa}, where we include the uncertainties coming
        from the DM profile slope in quadrature ($\Gamma=1.26\pm0.05$ at
        $3\sigma$); see  also Tab.~\ref{tab:cross-sections}.  The
        \textit{error-bars} indicate the annihilation cross-section preferred
        for the values $\Gamma=1.04$--$1.24$, as found from the GC  
        analysis in Ref.~\cite{Daylan:2014rsa}.
    \label{fig:cross-sections}}
\end{figure}

The energy spectrum of the excess emission as derived from the inner Galaxy
analysis of Ref.~\cite{Daylan:2014rsa} can be well fitted with secondary
photons from DM annihilation into hadronic final states.  Fits with the
harder photon spectra from annihilation into leptonic final states (caused by
final state radiation and, in the case of $\tau$ leptons, the decay of energetic 
neutral pions) are however disfavoured
on purely statistical grounds with high significance.  In case of annihilation
into $\tau^+\tau^-$ final states, this discrepancy can be alleviated by adding a
$\bar{b}b$ component of at least $20\%$ (in which case the resulting $\Delta\chi^2$ is still
worse by $\sim20$ than a pure $\bar b b$ fit).  An alternative can be additional photons 
from bremsstrahlung and Inverse Compton scattering~\cite{Lacroix:2014eea} --  though a 
sizeable effect requires very large branching ratios into $e^\pm$ or $\mu^\pm$ final
states, as e.g.~in the case of democratic annihilation to all leptons.

In Fig.~\ref{fig:cross-sections}, we show the values of the DM annihilation
cross-section and DM mass that are consistent with the GeV excess, based on the
findings of Ref.~\cite{Daylan:2014rsa}.  We increased the size of the
confidence regions to include the uncertainty in the DM profile slope --
$\Delta\Gamma\simeq0.05$ as
inferred from the inner Galaxy analysis -- in quadrature; this translates into
a relative uncertainty of $\Delta \langle\sigma v\rangle / \langle\sigma
v\rangle\simeq22$\,\%. These
numbers are also summarized in Tab.~\ref{tab:cross-sections} for convenience,
and will be used as benchmarks in our subsequent study of implications for
charged cosmic rays and radio emission.  Furthermore, the vertical error bars
indicate the range that is preferred by the GC analysis~\cite{Daylan:2014rsa},
which is in general higher than the range inferred from the inner Galaxy analysis.

\begin{table}
    \begin{tabular}{ccccc}
        \toprule
        Channel && Mass $m_\chi$ && Cross-section $\langle\sigma v\rangle$\\
        && [GeV] && [$10^{-26}$cm$^3$s$^{-1}$] \\\colrule
        $\bar{b}b$ && $35.5 \pm 4.2 $ &&     $ 1.7 \pm 0.3$ \\
        $\bar{c}c$ && $27.0 \pm 3.3 $ &&     $ 1.2 \pm 0.22$ \\
        $\bar{q}q$ && $18.5 \pm 2.1 $ &&     $ 0.72 \pm 0.13$ \\
        $\tau^+\tau^- (80\%)$ 
        && $9.3 \pm 0.8 $ &&     $ 0.7 \pm 0.12$ \\
        $\rm mass$ && $33.3 \pm 3.9 $ &&     $ 1.6 \pm 0.29$ \\
        dem && $21.9 \pm 3.1 $ &&     $ 1.1 \pm 0.2$ \\
        $\bar{s}s$ && $21.4 \pm 2.9 $ &&     $ 0.93 \pm 0.16$ \\
        $\rm charge$ && $18.7 \pm 2.3 $ &&     $ 1.3 \pm 0.23$ \\
        \botrule
    \end{tabular}
    \caption{List of benchmark annihilation channels that we consider in this
        work.
        Annihilation rates refer to a
        generalized NFW profile with central values $\Gamma=1.26$ and local
        density $\rho_\odot=0.3\rm\ GeV/cm^3$, using results from
        Ref.~\cite{Daylan:2014rsa} (see there for a definition of final states).  The 
        errors are from the statistical fit (95\% CL) and include additional uncertainties in
        $\Gamma=1.26\pm0.05$ ($3\sigma$ CL) in quadrature.  Taking into account
        the larger uncertainties in the slope $\Gamma$ as inferred from the GC
        analysis (see Fig.~\ref{fig:gamma-dependence})
        can furthermore change the values as
        indicated in Fig.~\ref{fig:cross-sections}; for $\Gamma=1.04$, e.g., the best-fit 
        value of $\langle\sigma v\rangle$ given in the table must be multiplied by 3.3.}
    \label{tab:cross-sections}
\end{table}

\subsection{Astrophysical scenarios}

For completeness, we
will here briefly sketch astrophysical scenarios that might account for
the excess emission seen at the GC.  The arguably most plausible
explanation for at least part of the observed excess emission at and close to
the GC is the emission from a large number ($\sim1000$) of MSPs
below the point-source sensitivity of \Fermi\ LAT~\cite{Abazajian:2010zy}.  Up
to now, more than 40 MSPs have been observed in gamma rays by the \Fermi\
LAT~\cite{TheFermi-LAT:2013ssa}, with spectra that are compatible with the
spectrum of the extended source at the GC~\cite{Gordon:2013vta}
(unless the spectrum of the GC excess below 1 GeV is confirmed to be extremely
hard~\cite{Daylan:2014rsa}).  MSPs remain gamma-ray emitters
for billions of years, and it was argued that with kick velocities of the order
of $\sim 40\rm\ km/s$ they have the right properties to in principle account
for the steepness as well as the extension of the observed gamma-ray
excess~\cite{Hooper:2011ti}.  The main argument against a significant
contribution of MSPs to the GC excess is that it appears to be
non-trivial to find plausible source distributions that completely remain below
the \Fermi\ LAT point source {threshold~\cite{Hooper:2013nhl, Calore:2014oga,
Yuan:2014rca}}, while still being
compatible with the emission properties of the pulsar population that is
observed locally.

\medskip

The emission of TeV gamma rays in the Galactic ridge region as observed by
H.E.S.S.~(in the inner $|b|<0.3^\circ$ and $|\ell|<0.8^\circ$) is well
correlated with the distribution of molecular clouds that are observed in the
inner 200 pc around the GC by means of radio
observations~\cite{Aharonian:2006au}.  This strongly suggests that the diffuse
TeV gamma-ray emission is due to a hard population of cosmic-ray electrons or
protons, producing gamma rays either via bremsstrahlung or proton--proton
interactions.  It is plausible that the same populations
also contribute to the GC emission at GeV energies.  In the context of cosmic-ray
electrons, Ref.~\cite{YusefZadeh:2012nh} showed that the bremsstrahlung from an
electron population compatible with the observed synchrotron emission at the
GC could indeed produce the characteristic peaked GeV excess
emission.  The main argument against the interpretation in terms of cosmic rays
is the apparent extension of the GeV excess to $\sim$kpc distances from the
GC as well as its spherical symmetry, which does not resemble the
distribution of detected gas (see \textit{e.g.}~Ref.~\cite{Abazajian:2014fta}).
Recent counter examples which go beyond the typical assumptions of static
cosmic-ray equilibrium at the Galactic center were presented in
Refs.~\cite{Petrovic:2014uda, Carlson:2014cwa}. In both papers, the authors
discuss recent burst events (up to about one million years ago) that injected
either high-energy electrons~\cite{Petrovic:2014uda} or
protons~\cite{Carlson:2014cwa} at the Galactic center, giving after a diffusion
period rise to a quasi-spherical excess emission around the GC.

\section{Constraints from other indirect detection channels}
\label{sec:constraints}

We now turn to a discussion of other messengers for the indirect detection of DM 
than gamma rays. In all these cases the source function is given by the
differential injection rate of particles from DM annihilation,
$Q(r,E)=\frac12\langle\sigma v\rangle dN/dE\, (\rho_\chi/m_\chi)^2$, that was already 
introduced in Eq.~(\ref{eq:qdef}) in the context of gamma rays. As stressed in the 
previous Section, \textit{cf.}~Figs.~\ref{fig:gamma-dependence} and 
\ref{fig:cross-sections}, this quantity is rather tightly constrained if the GeV excess is 
indeed explained by DM. In consequence, the intrinsically large uncertainties due to 
the DM distribution that limits from indirect detection are typically hampered by are 
greatly reduced in our case. In particular, they do not depend on the overall 
normalization of the DM density profile (conventionally expressed in terms of the 
local DM density $\rho_\odot$).

The spectrum of the DM signal is determined by $dN/dE$, i.e.~the differential number of 
a given species of cosmic-ray particles that are produced per annihilation. We
obtain these functions from \texttt{DarkSUSY 5.1.1}~\cite{DS}, which  provides tabulated 
fragmentation functions for various possible annihilation channels based 
on the event generator \texttt{PYTHIA 6.414}~\cite{Sjostrand:2006za} (for light 
quarks $q=u,d,s$, we take instead the spectra provided in Ref.~\cite{Cirelli:2010xx} 
as those are currently not implemented in  \texttt{DarkSUSY}).

\subsection{Antiprotons}

Final state quarks from DM annihilation in the Galaxy will fragment and produce 
antiprotons  \cite{Silk:1984zy}. Unlike gamma rays, those are deflected by 
stochastically distributed inhomogeneities in the galactic magnetic field such that the 
resulting propagation  can be  modelled by a diffusion process \cite{Salati:2010rc}.  
On the other hand, there are no primary but only secondary sources of 
astrophysical antiprotons:  these are produced through the collisions  of cosmic rays, in 
particular protons, with the interstellar medium. This astrophysical background is 
extremely well understood and can nicely be described in relatively simple 
semi-analytical diffusion models with cylindrical symmetry 
\cite{Donato:2001ms,Bringmann:2006im}. Fitting the parameters of those models to 
{\it other} cosmic-ray data, in particular other observed secondary to primary ratios 
like the boron over carbon ratio B/C \cite{Maurin:2001sj}, results in a prediction for 
the antiproton background that  is both tightly constrained and provides a very good 
fit to the data.

The main uncertainty in the background prediction derives from  the range of 
propagation parameters compatible with B/C and from uncertainties in the nuclear 
cross sections for the production of antiprotons. For the energy range we are 
interested in here, both effects can independently affect the flux by up to 
about 30\%~\cite{Salati:2010rc}. For recent studies that find similar values, and 
offer more detailed discussions about the underlying systematics, see 
Refs.~\cite{diMauro:2014zea,Kappl:2014hha}. 
In our analysis, we will take into account the full 
range of uncertainty in 
the background prediction by two independent parameters 
$\alpha_\mathrm{prop},\alpha_\mathrm{nuc}\in[0,1]$ that interpolate linearly between 
the minimal and maximal predictions for the secondary flux due to these two effects 
(for which we use the results from Refs.~\cite{Bringmann:2006im} and 
\cite{Donato:2001ms}, respectively). Low-energy antiprotons are furthermore 
affected by local effects like adiabatic energy losses in the expanding solar wind and 
diffusion in the solar magnetic field, often collectively referred to
as  solar modulation (see \textit{e.g.}~Ref.~\cite{Fornengo:2013xda} for a recent 
discussion). 
For the energies of interest to our analysis this is extremely well described 
by the force-field approximation \cite{Gleeson:1968zza,1987AA184119P}, see 
discussion in section~\ref{sec:antiprotons_discussion}, implying that a single 
parameter -- the Fisk potential $\phi_F$ -- is sufficient to relate the local
interstellar (LIS)
antiproton flux to the one measured at the top of the atmosphere (TOA).

\begin{figure}
    \begin{center}
        \includegraphics[width=\linewidth]{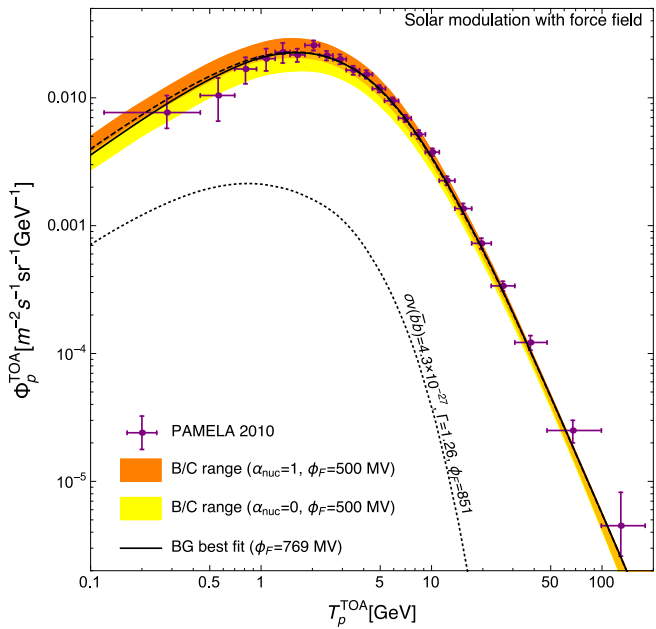}
    \end{center}
    \caption{PAMELA antiproton data \cite{Adriani:2012paa} as measured on top of 
    the atmosphere (TOA). The coloured bands show the prediction for the 
    astrophysical background (BG), with the width of each band deriving from 
    uncertainties in the propagation parameters left from the B/C analysis. The two 
    different bands bracket the uncertainty from nuclear cross sections, where  the 
    maximal (minimal) flux corresponds to the analysis performed in 
    Ref.~\cite{Donato:2001ms} (\cite{Bringmann:2006im}). The best-fit BG model is 
    given by the solid line. For comparison, the dotted and dashed lines also show the case of a fiducial WIMP 
    with mass 34 GeV, annihilating to $\bar bb$ with a rate barely allowed at 95\%CL 
    (see Fig.~\ref{fig:pbarlimits}).}
    \label{fig:pbarfluxes}
\end{figure}

For our analysis we use the newest release of PAMELA data that result from 
measurements between June 2006 and January 2010 \cite{Adriani:2012paa}, 
featuring significantly reduced error bars with respect to any previous antiproton 
data.\footnote{Note that the error bars stated in that data release are statistical
only. Systematical error bars are expected to be of the same order as in the
first release \cite{Adriani:2010rc} of PAMELA data \cite{privateComm}. In our analysis, we 
thus add those in quadrature.}  
These data agree remarkably well with the much older predictions 
\cite{Bringmann:2006im} for the antiproton background we are testing against: with 
the three free parameters described above, we find $\chi^2/d.o.f.=10.1/(23-3)=0.51$ 
for the best fit point. Clearly, this provides an important test for the underlying 
diffusion model. In Fig.~\ref{fig:pbarfluxes} we plot this best-fit prediction for the 
background as a solid line, along with the PAMELA data points. The yellow band in 
that figure corresponds to a choice of nuclear cross section parameterization that 
minimizes the flux (as adopted in Ref.~\cite{Bringmann:2006im}, this corresponds to 
setting $\alpha_\mathrm{nuc}=0$ in our analysis), while the orange band 
corresponds to a choice that maximizes the flux (as in Ref.~\cite{Donato:2001ms}, 
corresponding to our $\alpha_\mathrm{nuc}=1$). In both cases, the width of these 
bands is given by the uncertainty in the propagation parameters that results from 
the B/C analysis (which corresponds to varying our parameter 
$\alpha_\mathrm{prop}$ from 0 to 1).

The contribution to the antiproton flux from DM annihilation \cite{Silk:1984zy} is 
subject to much larger theoretical uncertainties than what is illustrated by the 
coloured bands in Fig.~\ref{fig:pbarfluxes} for the astrophysical background 
\cite{Donato:2003xg}. The main  reason for this is that DM annihilation is very 
efficient in a rather large part of the halo, implying that it probes a much larger 
volume of the diffusion zone than the B/C analysis that is restricted to sources in the 
Galactic disk. In particular, the antiproton flux from DM is mostly sensitive to the 
thickness $L$ of the diffusion zone perpendicular to the Galactic plane, while B/C 
essentially only constrains the {\it ratio} of $L$ and the diffusion coefficient $D$ 
\cite{DiBernardo:2009ku}. 
While the B/C analysis in principle allows a diffusion zone as small as 
$L\sim1$\,kpc, a vertical extension of $L\sim10$\,kpc is preferred when taking into 
account radioactive isotopes \cite{Putze:2010zn}, with similar results obtained when 
adding  gamma rays \cite{Timur:2011vv, FermiLAT:2012aa} and cosmic-ray electrons 
\cite{Delahaye:2008ua,Lavalle:2010sf} to the analysis. Also radio  
\cite{Bringmann:2011py,DiBernardo:2012zu} and low-energy cosmic-ray positron
\cite{Lavalle:2014kca} data have been shown to be clearly inconsistent with a halo 
size as small as  $\sim1$\,kpc. With this in mind, we will in the 
following mainly use the recommended reference model, 'KRA', of the recent 
comprehensive analysis presented in Ref.~\cite{Evoli:2011id}, which features 
$L=4$\,kpc (and is very similar to the best-fit model of 
Ref.~\cite{DiBernardo:2009ku}). For the propagation of primary antiprotons we 
use \texttt{DarkSUSY}~\cite{DS}, which implements the semi-analytical solution
of the diffusion equation described in Refs.~\cite{Bergstrom:1999jc,Evoli:2011id}.

We use the likelihood ratio test \cite{Rolke:2004mj} to determine limits on a possible 
DM contribution to the antiproton flux measured by PAMELA. For the  likelihood 
function, we adopt a product of normal distributions over each data bin $i$,
\be
\mathcal{L}=\Pi_i\,N(f_i|\mu_i,\sigma_i)\,,
\ee
where $f_i$ is the measured value, $\mu_i$ the total antiproton flux predicted by the 
model and $\sigma_i$ its variance. For a given mass and annihilation channel, the 
DM contribution enters with a single degree of freedom that parameterizes the 
non-negative signal normalization (and which we will always express in terms of the 
annihilation rate). 95\%CL  upper limits on $\langle \sigma v\rangle$ are thus 
derived by increasing the signal normalization from its best-fit value until 
$-2 \ln \mathcal{L}$ has changed by 2.71, while re-fitting ('profiling over') the 
parameters $(\alpha_\mathrm{prop},\alpha_\mathrm{nuc},\phi_F)$ of the 
background model. 

\begin{figure}
    \begin{center}
        \includegraphics[width=\linewidth]{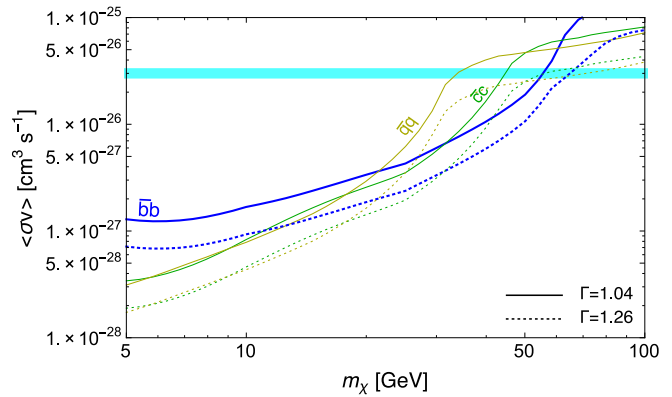}
    \end{center}
    \caption{Limits on the annihilation rate of DM into quark final states from our 
    analysis of the PAMELA antiproton data. Solid lines refer to the generalized NFW 
    profile of Eq.~(\ref{nfw}) with $\Gamma=1.04$ and are essentially 
    indistinguishable from the standard NFW ($\Gamma=1$) case; dotted lines show 
    the case for $\Gamma=1.26$.}
    \label{fig:pbarlimits}
\end{figure}

In Fig.~\ref{fig:pbarlimits}, we show the resulting limits on $\langle \sigma v\rangle$ 
as a function of the DM mass $m_\chi$, for all quark final states and two 
representative values of the $\Gamma$-parameter of the generalized NFW profile 
of Eq.~(\ref{nfw}). Limits for the standard NFW profile ($\Gamma=1$) are essentially 
indistinguishable from the $\Gamma=1.04$ case displayed here. These limits are 
one of our main results and rather strong, excluding the cross section 
$\langle\sigma v\rangle_{\rm therm}\equiv3\cdot10^{-26}$cm$^3$s$^{-1}$ typically 
favoured by thermally produced DM up to 
masses of $m_\chi\sim 35-55$\,GeV for an NFW profile (depending on the channel). 
There are two main reasons why we could improve previous limits 
\cite{Evoli:2011id,Kappl:2011jw,Cirelli:2013hv,Fornengo:2013xda} by a factor of roughly 2--5 at the 
DM masses of interest here (while the limits presented in Ref.~\cite{Tavakoli:2013zva} 
are actually slightly {\it stronger} than ours):
 i) we use the only recently published update of 
PAMELA data \cite{Adriani:2012paa} rather than the first public release 
\cite{Adriani:2010rc} and ii) we employ an improved statistical treatment of the 
background uncertainties (see Section~\ref{sec:antiprotons_discussion} for a more 
detailed discussion).\footnote{
Below $m_\chi\sim50$\,GeV, the limits presented in Ref.~\cite{Cirelli:2013hv} become
furthermore significantly weaker due to the deliberate choice of not including data with
$T<10$\,GeV.
}
When 
comparing these results to Fig.~\ref{fig:cross-sections}, we find that any 
interpretation of the gamma-ray excess as being due to DM annihilating into quark 
final states is in strong tension with the cosmic-ray antiproton data.

\begin{figure}
    \begin{center}
        \includegraphics[width=\linewidth]{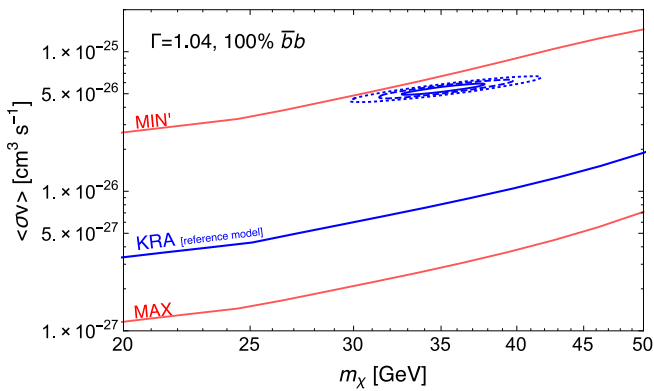}
    \end{center}
    \caption{Reference $\bar{p}$ limits (thick line) and effects of varying the propagation 
    scenario for $\bar bb$ final states. As in Fig.~\ref{fig:pbarlimits}, the area above the 
    lines is excluded at 95\%CL. For comparison, also the signal region for a DM 
    interpretation of the 
    gamma-ray excess in the inner Galaxy \cite{Daylan:2014rsa} is plotted, rescaled
    to $\Gamma=1.04$.}
    \label{fig:pbar_limits_prop}
\end{figure}

Let us, finally, comment on the impact of different propagation scenarios on our limits. 
Conventionally, the corresponding uncertainty is bracketed by two sets of propagation 
parameters, 'MIN' and 'MAX', that are consistent with the B/C analysis and, respectively, 
minimize and maximize the primary antiproton flux from DM annihilation 
\cite{Donato:2003xg}. As we have stressed before, however, there are several additional 
observations that constrain these parameters much better than the B/C analysis alone, 
such that the range of allowed fluxes spanned by MIN and MAX must be considered 
unrealistically large. In order to give a conservative indication of the involved 
astrophysical uncertainties, 
and in order to follow the typically adopted procedure, we still show in 
Fig.~\ref{fig:pbar_limits_prop} how our 
limits change when varying the propagation parameters within these ranges.\footnote{
Given that $L=1$\,kpc as featured by the MIN model proposed in Ref.~\cite{Donato:2003xg} 
has in the meantime been  firmly 
ruled out, however, we used instead a MIN' model with the same parameters as MIN but 
with $L=2$\,kpc and a diffusion coefficient of $D_0=9.65\cdot10^{26}$\,cm$^2$s$^{-1}$. 
This takes into account the lower bound of $L\geq2$\,kpc from radio observations 
\cite{Bringmann:2011py,DiBernardo:2012zu} and the fact that B/C only is sensitive to 
$L/D_0$ \cite{Maurin:2001sj,DiBernardo:2009ku}. Note that even $L=2$\,kpc is very
conservative in light of the recent analysis of low-energy positron data \cite{Lavalle:2014kca}.
} 
As can be seen from this figure, the DM interpretation of the excess becomes compatible 
with limits from the PAMELA antiproton data only in the most unfavourable 
case of propagation parameters -- at least within the cylindrical two-zone diffusion model 
that is commonly considered. Antiproton data from the AMS-02 experiment on board of 
the international space station  may improve limits on a DM contribution by as much as one
order of magnitude with respect to the current PAMELA data \cite{Evoli:2011id,Cirelli:2013hv,
Fornengo:2013xda}. Expected to be 
published in less than a year from now, AMS-02 data will thus either show an excess also in 
antiprotons or allow to rule out the DM hypothesis with rather high confidence. Similar 
conclusions apply more generally to other quark annihilation channels and DM profiles 
than what is shown explicitly in Fig.~\ref{fig:pbar_limits_prop} 
(i.e.~$\bar bb$ and $\Gamma=1.04$).

\subsection{Positrons}

\begin{figure}
    \begin{center}
        \includegraphics[width=\linewidth]{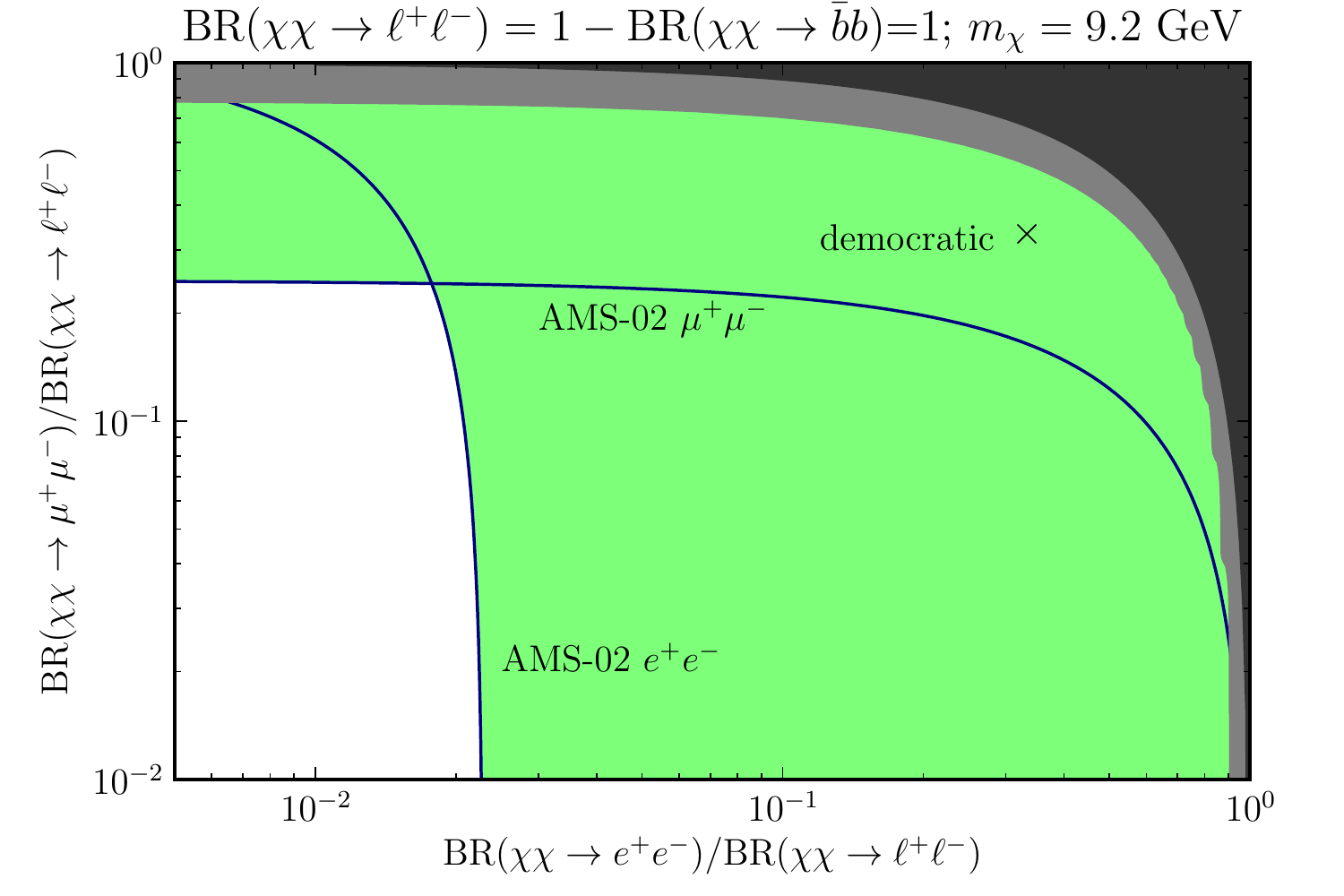}
    \end{center}
    \caption{Upper limits (95\% CL) on the relative branching ratios into
        leptonic two-body final states, as derived from a spectral analysis 
        \cite{Bergstrom:2013jra} of AMS-02 positrons.  We assume $100\%$
        annihilation into leptonic final states.  For each point, we determine
        the DM mass and cross-section by a fit to the gamma-ray spectrum 
        of the inner Galaxy excess \cite{Daylan:2014rsa}, assuming
        $\Gamma=1.26$.  The green regions are
        excluded, while the gray region shows where the spectral fit to the
        GeV excess worsens significantly (see text for details). The white 
        area shows the remaining allowed parameter space, corresponding 
        to an almost pure $\tau^+\tau^-$ final state. Note, however, that this
        gives a fit to the data that is still much worse (by about $\Delta
        \chi^2\sim130$) than a fit with a $\bar{b}b$ final state.}
    \label{fig:triangle}
\end{figure}

\begin{figure*}
    \begin{center}
        \includegraphics[width=0.49\linewidth]{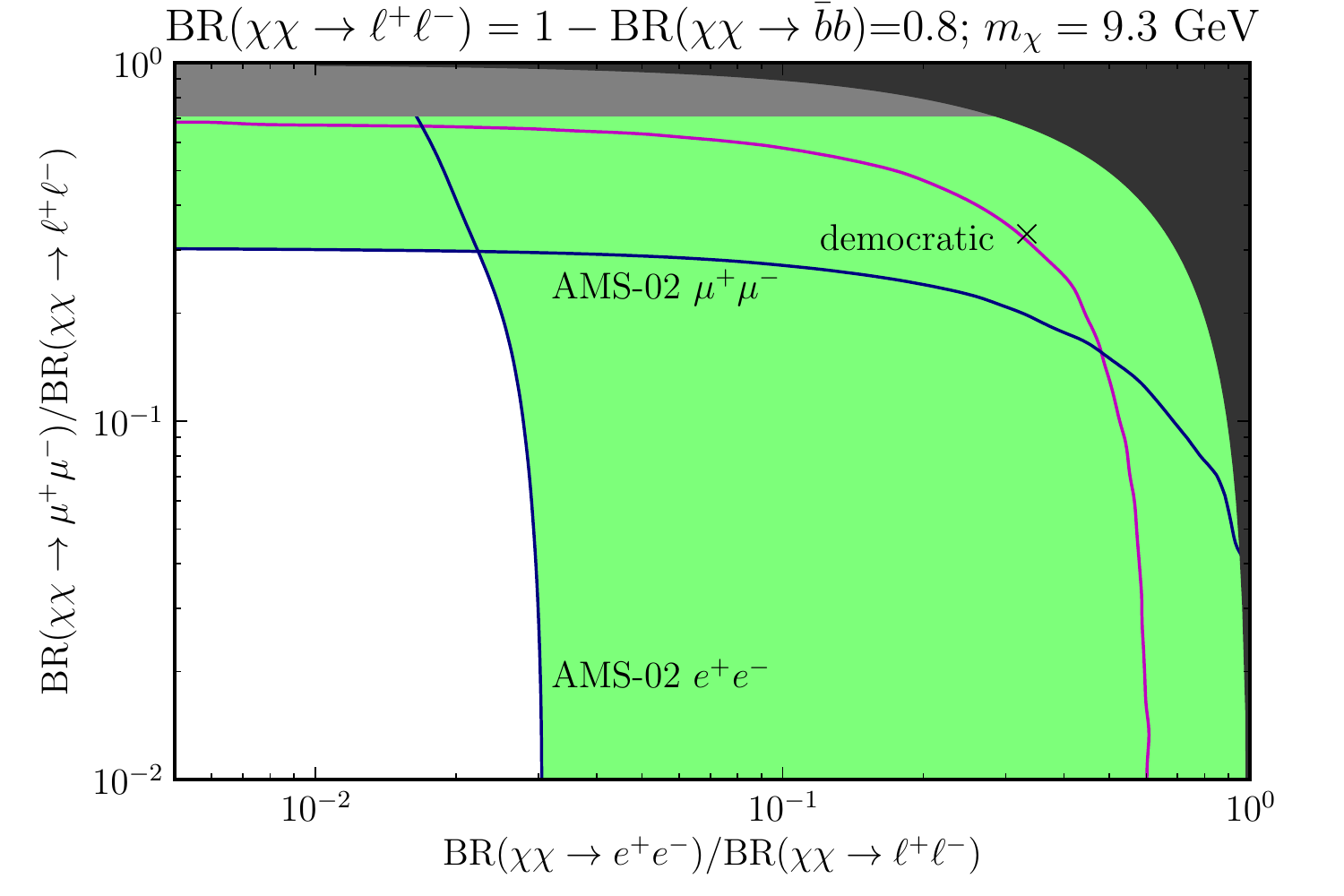}
        \includegraphics[width=0.49\linewidth]{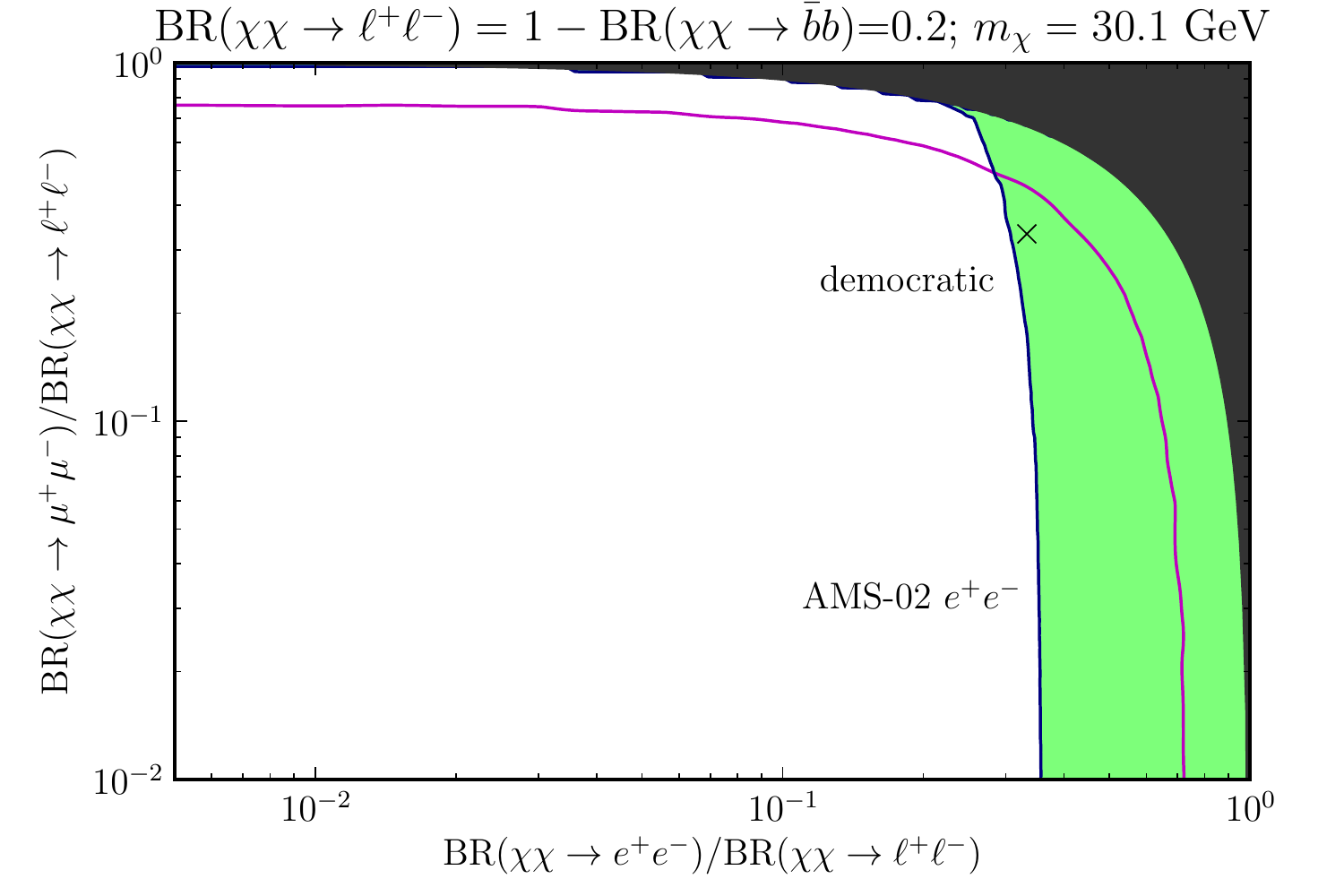}
    \end{center}
    \caption{Same as Fig.~\ref{fig:triangle}, but for non-zero branching
        ratios into $\bar{b}b$ final states. The magenta line in the left
        (right) panel indicates where the best-fit DM mass exceeds 14 (34)
        GeV.}
    \label{fig:triangle2}
\end{figure*}

The energy spectrum of cosmic-ray positrons as well as the \textit{positron
fraction} (the fraction of positrons in the total electron and positron flux)
was recently measured with unprecedented precision by the
AMS-02~\cite{Aguilar:2013qda} experiment, in the energy range 0.5 to 350 GeV.
AMS-02 confirmed the rise in the positron fraction at energies above 10 GeV
that was previously observed by PAMELA~\cite{Adriani:2008zr} and Fermi
LAT~\cite{FermiLAT:2011ab}, but with significantly smaller statistical and
systematical errors.  This allowed for the first time a dedicated spectral
search for signals from light ($m_\chi\lesssim 350\rm \ GeV$) DM particles 
annihilating into leptonic final
states~\cite{Bergstrom:2013jra, Ibarra:2013zia}, in a way that is largely
independent of the origin of the rise in the positron fraction itself.  For DM
masses around 10 GeV, the limits on the annihilation cross-section into
$e^+e^-$ ($\mu^+\mu^-$) are very tight and around $1.2\times10^{-28}\rm\ cm^3
s^{-1}$ ($1.3\times10^{-27}\rm\ cm^3 s^{-1}$)~\cite{Bergstrom:2013jra}.

The AMS-02 measurements of the positron fraction have important consequences
for the DM interpretation of the GeV excess.  We will here consider the option
that the GeV excess is dominantly caused by annihilation into leptonic two-body
final states, with a possible admixture of $\bar{b}b$.  This scenario is
described by the branching ratios into the three charged lepton families ($e^+e^-$,
$\mu^+\mu^-$ and $\tau^+\tau^-$ final states) as well as $\bar{b}b$.  For a
given set of branching ratios, we calculate the prompt gamma-ray emission using
\texttt{DarkSUSY}~\cite{DS}.
We refit the energy spectrum of the GC excess emission (Fig.~5 in
Ref.~\cite{Daylan:2014rsa}) to obtain the total annihilation cross-section and
DM mass, assuming a DM profile with $\Gamma=1.26$.  During the fit, we
constrain the DM mass to be larger than $9.2\rm\ GeV$ to approximately account
for the fact that bremsstrahlung and inverse Compton emission can potentially
contribute at low energies to reconcile the spectra of mixed leptonic final
states with the measurements~\cite{Lacroix:2014eea} (though this arguments only
works in case of sizeable branching ratios to $e^+e^-$ final states, as is the
case for the democratic scenario that Ref.~\cite{Lacroix:2014eea} considers).

For a given set of branching ratios and the implied DM mass and total
cross-section, we adopt the AMS-02 limits from Ref.~\cite{Bergstrom:2013jra} to
decide whether a scenario is excluded. We use here the central values of the
limits from Ref.~\cite{Bergstrom:2013jra}; uncertainties in the local radiation
field allow these limits to weaken by maximally a factor of two.  We use a
reference value of $\Gamma=1.26$ throughout. Note that a shallower profile
would strengthen the AMS-02 limits, which mostly depend on the
annihilation rate at the Sun's position, by a factor of $\sim3$; variations in the scale radius $r_s$
allow for an additional factor two up or down in the annihilation rate
(\textit{cp.}~Fig.~\ref{fig:gamma-dependence}).

The results of this procedure are shown in Figs.~\ref{fig:triangle}
and~\ref{fig:triangle2}:  In Fig.~\ref{fig:triangle} we consider the purely
leptonic case, i.e.~${\rm BR}(\chi\chi\to \bar{b}b)=0$.  In this figure, the best-fit 
DM mass always stays close to the imposed lower limit of 9.2 GeV. The white areas 
are allowed and the green areas excluded by limits on $e^+e^-$ or $\mu^+\mu^-$ 
final states at $95\%$ CL; the
gray area indicates where the formal fit to the data becomes worse by 
$\Delta \chi^2 \gtrsim 25$.
In Fig.~\ref{fig:triangle2} we show the same situation, but assume
that $20\%$ (left panel) or $80\%$ (right panel) go into $\bar{b}b$ final
states.  In each case, we indicate in the figure header  the best-fit mass
that we obtain in the limit $\rm BR(\chi\chi\to e^+e^-) = BR(\chi\chi\to
\mu^+\mu^-)=0$.  As one can see from these plots, the democratic case, 
featuring equal leptonic branching ratios, is
clearly excluded from \mbox{AMS-02} cosmic-ray positron data. 

Let us stress again that leptons alone do not feature a spectral 
shape consistent with that of the GeV excess. This holds not only for the extremely 
hard spectrum associated with light lepton final states, but also for the slightly softer spectrum 
from tau leptons. Due to the strong \mbox{AMS-02} constraints on the contribution from 
$\mu^\pm$ and $e^\pm$ final states for ${\rm BR}(\chi\chi\to \bar{b}b)\lesssim0.2$, we find that 
this conclusion cannot be changed by including the effect of inverse Compton and 
bremsstrahlung processes, as suggested in Ref.~\cite{Lacroix:2014eea}. Note that this
holds even if annihilation is assumed to happen dominantly into $\mu^\pm$ and 
$\tau^\pm$, thereby evading the extremely strong constraints for $e^\pm$ final states:
In order to produce a reasonable fit to the data, one would in that case need
${\rm BR}(\chi\chi\to \mu^+\mu^-)\gg0.25$ \cite{Lacroix:2014eea}, which we identify 
in Fig.~\ref{fig:triangle} as being excluded by \mbox{AMS-02} data.

\subsection{Radio signals}
\label{subsec:radio}
Electrons and positrons from DM annihilation  (henceforth collectively referred to as 
electrons) are expected to emit synchrotron radiation when propagating through the 
Galactic magnetic fields. Here, we shall focus on corresponding radio signals from 
the GC. In particular, we will use the Jodrell Bank upper flux limit of 
50\,mJy from this region \cite{1976MNRAS.177..319D},
obtained for a frequency of 408\,MHz, to constrain the DM annihilation rate. As has been 
noticed before \cite{Gondolo:2000pn,Bertone:2001jv,Aloisio:2004hy,Regis:2008ij,
Bertone:2008xr,Bringmann:2009ca,Mambrini:2012ue,Laha:2012fg,Asano:2012zv},
the resulting constraints are typically rather strong.  In this
subsection we will derive our baseline constraints and defer a critical
discussion of the steps to Section~\ref{sec:radio_discussion}.

The arguably most critical -- yet, as we shall see, realistic -- assumption that 
enters our analysis is that the electrons in the 
GC region lose their energy essentially {\it in situ}, via synchrotron radiation, 
implying that both free-streaming and diffusion effects can be neglected. This is 
motivated by the fact that one expects a much larger turbulent magnetic field
at $\mathcal{O}(1\rm\,pc)$ distances from the 
GC \cite{1992ApJ...387L..25M} than the average Galactic value 
of $\sim\SI{6}{\micro G}$ \cite{Beck:2013bxa}. In order to get a 
quantitative idea of the magnetic field strength that is required, we will
here
assume that diffusion at $\lesssim 1\rm\ pc$ from the GC is well described by
Bohm diffusion (see Ref.~\cite{Bertone:2001jv} for a similar treatment); in
Section~\ref{sec:radio_discussion} we will argue that this assumption can in
fact be relaxed by several orders of magnitude.
In case of Bohm diffusion, the scattering length of the diffusion process is given by the gyroradius
$r_g$, leading to a diffusion constant \(D_\text{Bohm}=\frac13r_g c=E_ec/3eB\).
The 
length scale $l_\mathrm{diff}\simeq(D_\text{Bohm} t_\mathrm{loss})^{1/2}$ over which
relativistic electrons 
propagate during their synchrotron energy loss time, $t_\mathrm{loss}\approx
E/b(r,E)=3m_e^4c^7/2e^4B^2E$ where $b(r,E)$ is the loss rate, should then be 
significantly smaller than the DM density scale height 
$l_\chi\equiv|\rho_\chi(r)/\rho_\chi'(r)|$, i.e.~\cite{Asano:2012zv}
\be
\frac{l_\mathrm{diff}}{l_\chi}\simeq\frac{m_e^2c^4}{\sqrt2e^{5/2}l_\chi B^{3/2}}\lesssim1\,.
\ee
For the generalized NFW profile that we consider here, see Eq.~(\ref{nfw}), this
implies a lower limit on the magnetic field strength of
\be
\label{eq:Blimit}
    B(r)\gtrsim 4\,\Gamma^{2/3}\left(\frac{\SI{}{pc}}{r}\right)^{2/3}{\micro
    G}
\ee
for the diffusion of electrons from DM annihilation at the GC to be negligible. 
Once this condition is satisfied, the resulting limits
will actually {\it decrease} with increasing $B$ (while the opposite is true in the 
regime where diffusion cannot be neglected, see
\textit{e.g.}~Ref.~\cite{Boehm:2010kg}).  
\medskip

Observationally, the magnetic field in the Galaxy can be inferred only indirectly via the  
Faraday effect. The resulting rotation of polarized radio
waves with wavelength $\lambda$ is given by $\beta = \lambda^2\times{\rm RM}$, 
where the \emph{rotation measure} {\rm RM} is proportional to the integral over
the line-of-sight magnetic field $B(r)$ and the electron density $n(r)$, ${\rm RM}
\propto \int B(r) n(r)$.  

At the distances that are of interest for our radio discussion, $\sim0.1\rm\
pc$, Ref.~\cite{Eatough:2013nva} infers the magnetic field from 
multi-wavelength observations of the recently discovered magnetar PSR
J1745-2900, which has a rotation measure of 
${\rm RM}\sim 7\times 10^4$\,rad\,m$^{-2}$ at a projected distance of $0.12\rm\ pc$ 
from the central black
hole, Sagittarius A$^\ast$ (Sgr A$^\ast$).\footnote{
For comparison, the highest RM of {\it any} Galactic source, 
\mbox{$\sim5\times 10^5$\,rad\,m$^{-2}$}, is associated with the radio emission of Sgr A$^\ast$ 
itself~\cite{Marrone:2006vu}.
}
Together with an observed dispersion measure of $\sim 1.8\times 10^3 \rm\ cm^{-3}\,pc$, 
which determines the column density of electrons towards the pulsar, this allows 
to derive a very conservative lower limit on the magnetic field of $50\ \mu\rm G$ 
(assuming that all electrons along the line-of-sight are localized close to the
Galactic center, and that no turbulent field components and/or field reversals reduce
the RM).  A more realistic estimate 
gives a much larger lower limit of about 8 mG \cite{Eatough:2013nva}, but it is 
interesting to note that already the extremely conservative limit satisfies 
Eq.~(\ref{eq:Blimit}) if Bohm diffusion is realized.

\medskip

\begin{figure}[t]
     \begin{center}
         \includegraphics[width=\linewidth]{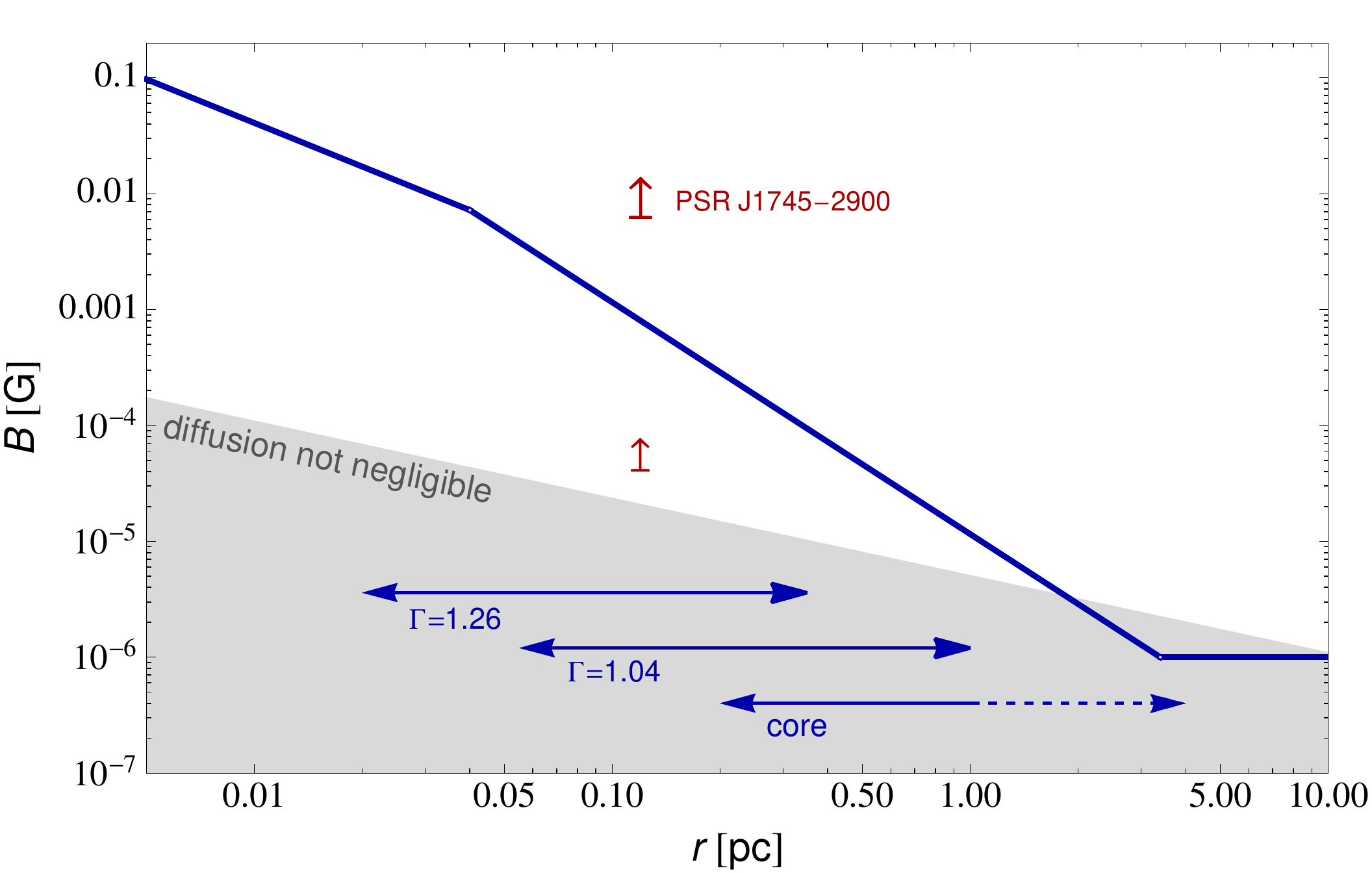}
     \end{center}
     \caption{{\it Solid line}: Simplified model for the magnetic field profile close to the 
     GC black hole \cite{Bertone:2008xr}, assuming energy equipartition inside the accretion 
     volume and magnetic flux conservation outside. The {\it gray area}  defines the 
     domain where Bohm diffusion can no longer be neglected (as assumed in our 
     analysis, \textit{cf.}~Eq.~(\ref{eq:Blimit})). Lower limits (in red) refer to ultra-conservative 
     and realistic field values, respectively, inferred from  multi-wavelength 
     observations of the recently discovered magnetar PSR J1745-2900 
     \cite{Eatough:2013nva}. Horizontal arrows indicate the ranges of galactocentric
     distances that, depending on the profile, contribute $\sim95\%$ of the
     annihilation signal flux at 408\,MHz in the 4'' cone observed in 
     Ref.~\cite{1976MNRAS.177..319D} (see text for details).
     For our actual limits, we conservatively take only radio fluxes 
     from the inner 1 pc around the GC into account.}
     \label{fig:Bfield} 
 \end{figure}

\begin{figure*}[t]
     \begin{center}
         \includegraphics[width=.494\linewidth]{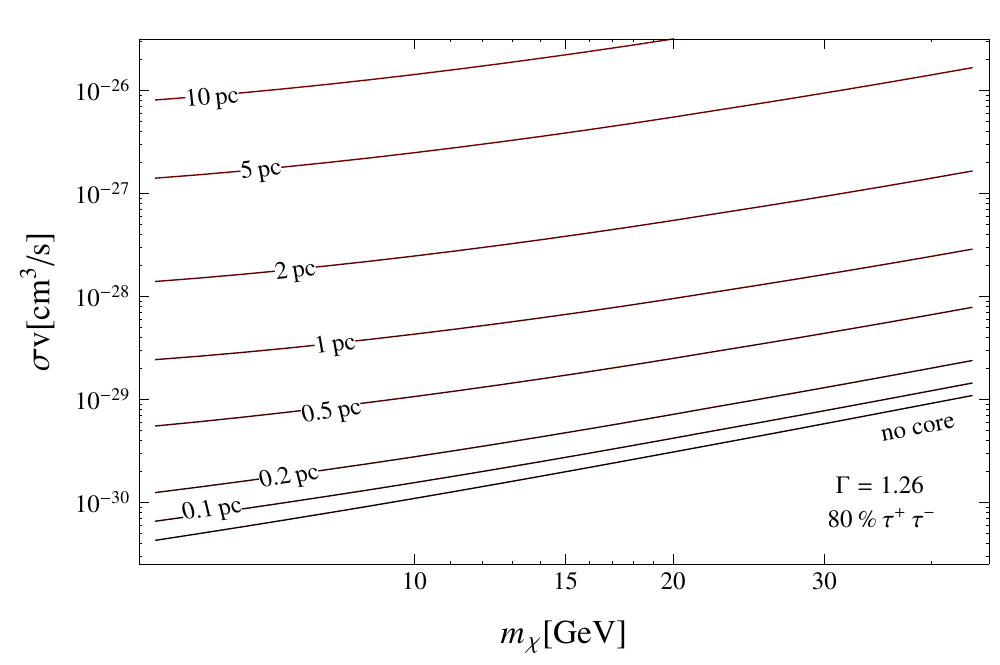}
         \includegraphics[width=.494\linewidth]{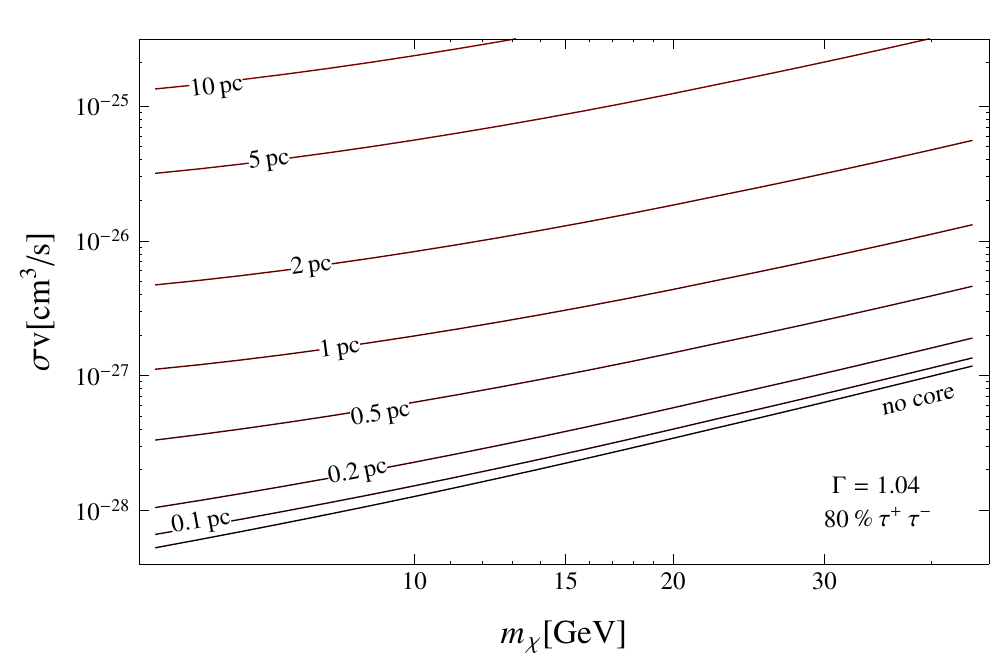}
     \end{center}
     \caption{{\it Left panel}: Limits from radio observations on the annihilation rate of 
     DM particles into 80\% $\tau^+\tau^-$ and 20\% $\bar bb$,  for a generalized 
     NFW profile with an inner slope of $\num{1.26}$ (lowest line). The other curves show  
     the same limits when adding an artificial core to the DM profile with a
     core size $r_c$ as indicated (i.e.~assuming a 
     constant profile for galactocentric distances smaller than what is stated next to the 
     respective curve).   {\it Right panel}: same as left 
     panel but with an inner slope of $\Gamma=1.04$.}
          \label{fig:radiotautau} 
\end{figure*}

For definiteness, we follow the often adopted assumption 
\cite{1992ApJ...387L..25M} that the magnetic field near the galactic center is mainly 
powered by the central SMBH. Concretely, this means that we assume an approximate 
equipartition of magnetic, kinetic and gravitational energy 
inside the accretion zone, i.e.~$B\propto r^{-5/4}$ for $r< R_\text{acc.}=\SI{.04}{pc}$ 
(see also Refs.~\cite{1971SvA....15..377S,1973ApJ...185...69S}). 
For  $r> R_\text{acc.}$, which is the region 
most relevant for our limits, magnetic flux conservation leads to $B\propto r^{-2}$. In 
Fig.~\ref{fig:Bfield}, we plot this magnetic field profile \cite{Bertone:2008xr} along with 
the condition given in  Eq.~(\ref{eq:Blimit}) and the observationally infered lower limits.
\medskip

If energy losses are dominated by synchrotron radiation, and the effect of diffusion 
can be neglected, the transport equation can be solved analytically. The total synchrotron 
flux density is then given by \cite{Bertone:2001jv,Bertone:2008xr,Asano:2012zv}
\begin{equation}
\label{eq:synflux}
F_\nu\simeq\frac{\langle\sigma v\rangle}{8\pi\nu R_\odot^2 m_\chi^2}\int E\rho_\chi^2(r)N_e(E)\der V\,,
\end{equation}
where $N_e(E)$ denotes the number of electrons (or positrons) per annihilation,  with 
energy larger than $E$.
In arriving at this expression, the monochromatic approximation for synchrotron radiation 
was used,
\be
\label{eq:Esynch}
E= \left(\frac{4 \pi m_e^3 \nu}{e B}\right)^\frac12=0.46\left(\frac{\nu}{{\rm GHz}}\right)^\frac12\left(\frac{B}{{\rm mG}}\right)^{-\frac12}
  \,{\rm GeV}\,,
\ee
which we checked affects our limits by less than 30\% for the masses of interest here.
The integration volume in Eq.~(\ref{eq:synflux}) is a cone corresponding to the $\num{4''}$ region 
 ($\sim\SI{.32}{pc}$ of diameter at the GC) 
 observed at Jodrell Bank \cite{1976MNRAS.177..319D}.  We restrict the integration 
 to a region $r\!<\!r_\mathrm{max}=1$\,pc where diffusion can be safely
 neglected, see Fig.~\ref{fig:Bfield}, 
  thus ignoring the synchrotron emission of electrons 
 created in regions where diffusion effects are not clearly negligible. While this restriction has 
 no significant effect on our limits for the case of a generalized NFW profile, it renders our 
 limits in the presence of a core, as discussed below, rather conservative. 
\medskip

In Fig.~\ref{fig:radiotautau} we show the results from confronting the DM hypothesis 
with the $\SI{408}{MHz}$ Jodrell Bank upper limit in the case where the annihilation 
of DM particles occurs with a branching ratio of 80\% into $\tau^+\tau^-$  and with 
20\% into $\bar bb$. Besides limits for a generalized NFW profile with 
$\Gamma=1.26$ (left panel) and $\Gamma=1.04$ (right panel), we also show limits 
for these profiles if an {\it ad hoc} cutoff  at a galactocentric distance $r_c$ is 
introduced in the DM density profile. Below this, the DM density is assumed to stay 
constant, i.e.~$\rho_\chi(r\!<\! r_c)=\rho_\chi(r_c)$ while 
$\rho_\chi(r\!>\! r_c)$ is given by Eq.~(\ref{nfw}). At much smaller scales 
than considered here, such a DM density plateau is expected to result from the large 
DM annihilation rate \cite{Berezinsky:1992mx}
(in extreme cases, also dynamical effects like the off-center formation of the SMBH 
\cite{Ullio:2001fb} or major SMBH merger events \cite{Merritt:2002vj} could significantly 
reduce the DM density at $r\lesssim1$\,pc, though this would not result in a complete 
flattening of the profile). Here, the postulated flattening at $r\!<\! r_c$
rather serves as a phenomenological parameterization of the maximal effect that 
uncertainties in the DM distribution at small scales may have on our limits. Let us 
stress that the DM interpretation of the GeV excess fixes the form of the density 
profile down to roughly 10 pc \cite{Daylan:2014rsa}, 
and that there is no particular reason to {\it expect} a cutoff at only slightly smaller scales.
GC radio observations thus place extremely tight constraints on annihilating DM for the 
steep density profiles considered here, at least if extending down to 
$r>r_c\sim1$\,pc.\footnote{
It is worth stressing that for such large core sizes, $r_c\gtrsim1$\,pc, the 
limits shown in Fig.~\ref{fig:radiotautau} are rather strongly affected by our conservative 
choice of restricting the volume for which we consider synchrotron integration. 
Changing the integration range in Eq.~(\ref{eq:synflux}) from 
$r\!<\!r_\mathrm{max}=1$\,pc to $r_\mathrm{max}=4$\,pc  inside the 4'' cone, for 
example, the constraints depicted for the $r_c=10$\,pc case would tighten 
by a factor of up to a few for $m_\chi=\mathcal{O}(10)\,\mathrm{GeV}$.
}
In fact, as we will discuss in more detail in
Section~\ref{sec:radio_discussion}, these limits generally depend much more strongly 
on the DM profile -- which is 
fixed once we accept the DM interpretation of the GeV excess -- than on the strength 
of the magnetic field. 

\begin{figure}[t]
     \begin{center}
         \includegraphics[width=\linewidth]{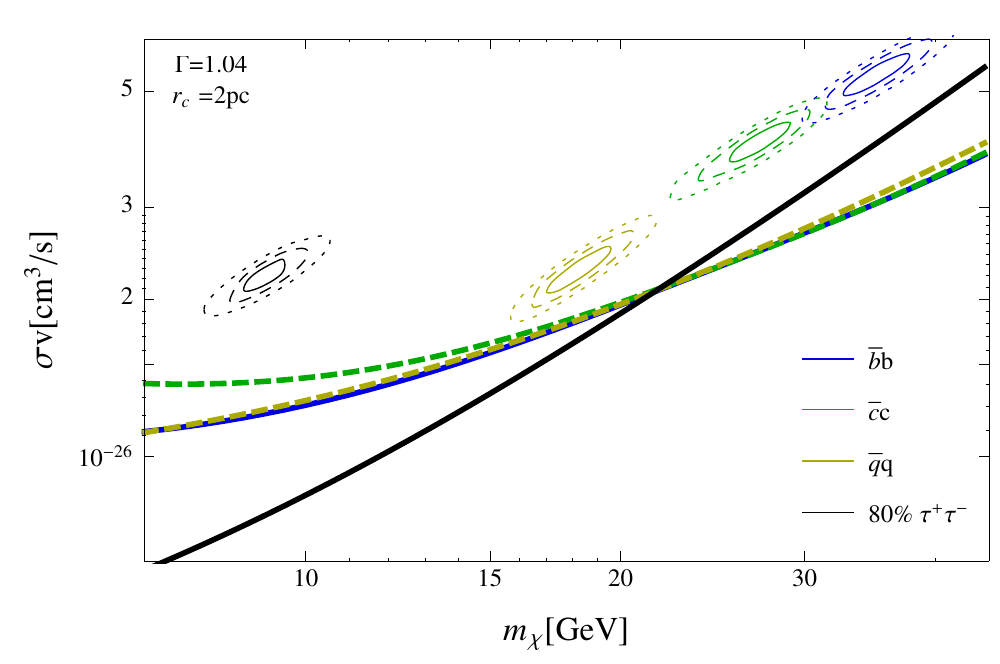}
     \end{center}
     \caption{Constraints from GC radio observations for various annihilation channels, 
     along with the corresponding contours characterizing the DM interpretation of  the 
     GeV excess in gamma rays \cite{Daylan:2014rsa}. These constraints assume an
     {\it ad hoc} core in the DM density profile at galactocentric distances smaller than
      $r_c=2$\,pc, i.e.~only slightly below the $\mathcal{O}(10)$\,pc 
     distance down to which the signal profile is observed in gamma rays (see Fig.~\ref{fig:radiotautau} for an 
     indication of how limits improve if the profile is assumed to continue to smaller 
     scales).}
     \label{fig:radioallchann} 
\end{figure}

In Fig.~\ref{fig:radioallchann}, we finally compare our radio limits directly to the gamma-ray 
signal claims \cite{Daylan:2014rsa} for various annihilation channels. Assuming that the 
observed profile extends down to a moderate $r_c\lesssim\SI{2}{pc}$, indeed, we find that {\it all} 
 channels are excluded as an explanation of the signal. Here, we 
used an inner slope of the DM profile of $\Gamma=1.04$; increasing this to $\Gamma=1.26$
would tighten the limits by roughly one order of magnitude for the DM masses that best
describe the excess.

For completeness, let us mention that a standard NFW profile \mbox{($\Gamma=1$)} without a
core, or a core at galacto-centric distances less than 0.1\,pc, leads to limits that are less 
than a factor of 3 weaker than for the case of $\Gamma=1.04$ considered above. This 
implies that for such a profile  thermal cross sections are excluded for DM masses below 
roughly 120 GeV (400 GeV) for $\tau^+\tau^-$ ($\bar bb$) final states. For an 
Einasto profile, 
on the other hand, we find constraints that are weaker by more than 2 orders of magnitude 
below $m_\chi\sim100$\,GeV, thus not probing thermal cross sections even for dark matter 
masses as small as a GeV. This large difference is easily understood by observing that for 
the small distance scales $r\sim1$\,pc that are most relevant in setting the limits, 
c.f.~Fig.~\ref{fig:Bfield}, the Einasto profile is already much shallower than the NFW profile.

\subsection{Further constraints}

Strong and robust constraints on light annihilating DM particles are in principle also 
provided by measurements of the {\it cosmic microwave background} 
\cite{Galli:2009zc,Slatyer:2009yq,Galli:2011rz,Slatyer:2012yq,Cline:2013fm,Lopez-Honorez:2013cua}.  
However, even projected limits from Planck data \cite{Cline:2013fm} for light lepton final 
states are much less constraining than the AMS-02 positron limits discussed above (and 
somewhat weaker for $\tau$ lepton final states). For quark final states, it will be possible 
to probe the thermal annihilation cross section up to masses of $\sim25$\,GeV. While 
this provides interesting and completely complementary limits compared to the ones 
derived from cosmic-ray antiproton observations, this only barely starts to constrain the 
assumed  GeV signal region for DM annihilation into light quarks 
(\textit{cf.}~Fig.~\ref{fig:cross-sections}).

\emph{Gamma-ray observations} of dwarf spheroidal galaxies of the Milky Way are
a further powerful and robust probe for the annihilation of DM.  As practically
background-free targets, and with a DM content that is well constrained from
observations of member stars, they were used by several groups to perform DM
searches both with space- and ground-based telescopes~\cite{Abdo:2010ex,
GeringerSameth:2011iw, Cholis:2012am, Ackermann:2013yva, Abramowski:2010aa,
Aliu:2012ga, Aleksic:2013xea}. For light DM, the most recent study was
based on a combined analysis of Fermi LAT observations of 15 dwarf spheroidal 
galaxies \cite{Ackermann:2013yva}.  Only upper limits
on the annihilation cross-section of DM could be found, and in case of
\textit{e.g.}~$\chi\chi\to\bar{b}b$ these are consistent with the DM interpretation of
the GeV excess at the Galactic center.  Interestingly, however, for DM masses around
$m_\chi=10$--$25\rm\ GeV$ the current analysis indicates a slight preference
for a signal at $\sim2.3\sigma$ (after the typical fluctuation level of the
extragalactic gamma-ray background has been taken into account).  Data
collected by the Fermi LAT over the upcoming years will help to sort out
whether this is a signal or merely a fluctuation.

Constraints from galaxy clusters~\cite{Ackermann:2010rg, Huang:2011xr,
Zimmer:2011vy, Ando:2012vu} or the extragalactic gamma-ray
background~\cite{Abdo:2010dk, Calore:2013yia}, are less stringent, since they
rely heavily on the distribution of substructures in DM halos.  However, a
cross-correlation between the distribution of DM in the local Universe and the
unresolved gamma-ray sky can be a promising venue to confirm the GeV excess at
higher latitudes~\cite{Camera:2012cj, Campbell:2013rua, Fornengo:2013rga,
Ando:2013xwa, Shirasaki:2014noa}. In passing, let us mention that previous constraints 
derived from GC observations \cite{Hooper:2012sr, Tavakoli:2013zva} are actually 
somewhat in tension with the observed GeV excess and its interpretation in terms of DM 
annihilation.

\emph{Radio searches} for DM annihilation have recently been performed also for 
other targets than the GC that we have discussed at length above.  The goal is 
usually not to identify a DM signal -- which is
extremely hard due to the large modeling uncertainties of signals and
backgrounds -- but to put upper limits on the DM annihilation rate. In
contrast to the constraints derived in the present work, which only depend weakly on 
the magnetic field, above a certain threshold, most other constraints critically depend on
the assumptions about magnetic field and cosmic-ray diffusion. 
Limits and prospective constraints from the inner Galaxy (a few
degrees off the GC) were discussed in Refs.~\cite{Crocker:2010gy,Fornengo:2011iq,
Mambrini:2012ue, Laha:2012fg,Wechakama:2012yb}.  The most recent analysis of Galaxy
clusters was presented in Ref.~\cite{Storm:2012ty}, leading to limits that are
stronger than corresponding cluster constraints from gamma-ray observations.
Radio searches in dwarf spheroidal galaxies suffer from the basically unknown
magnetic field in dwarfs~\cite{Natarajan:2013dsa, Spekkens:2013ik}. 

The Andromeda galaxy (M31),  on the other hand, has currently a lower star formation 
rate than the Milky Way, making it particularly suited for radio searches for DM. For 
realistic assumptions about the magnetic field, recent searches in M31 lead to very 
competitive constraints~\cite{Egorov:2013exa}. Depending also on the assumed halo 
model of M31, these are in some tension with the DM interpretation of the GeV excess.

\section{Discussion}
\label{sec:discussion}

In the previous Section we have seen that indirect searches using other messengers 
than gamma rays place very strong constraints on light annihilating DM, essentially 
for every possible fermionic annihilation channel (with the exception of neutrinos, 
which we have not discussed here). These constraints, if taken at face value,  have 
far-reaching implications for a possible DM interpretation of the GeV gamma-ray 
excess. In this Section, we will therefore re-assess their validity  and discuss
 systematic uncertainties not addressed so far.

\subsection{Antiprotons}
\label{sec:antiprotons_discussion}

\begin{figure}[t]
    \begin{center}
        \includegraphics[width=\linewidth]{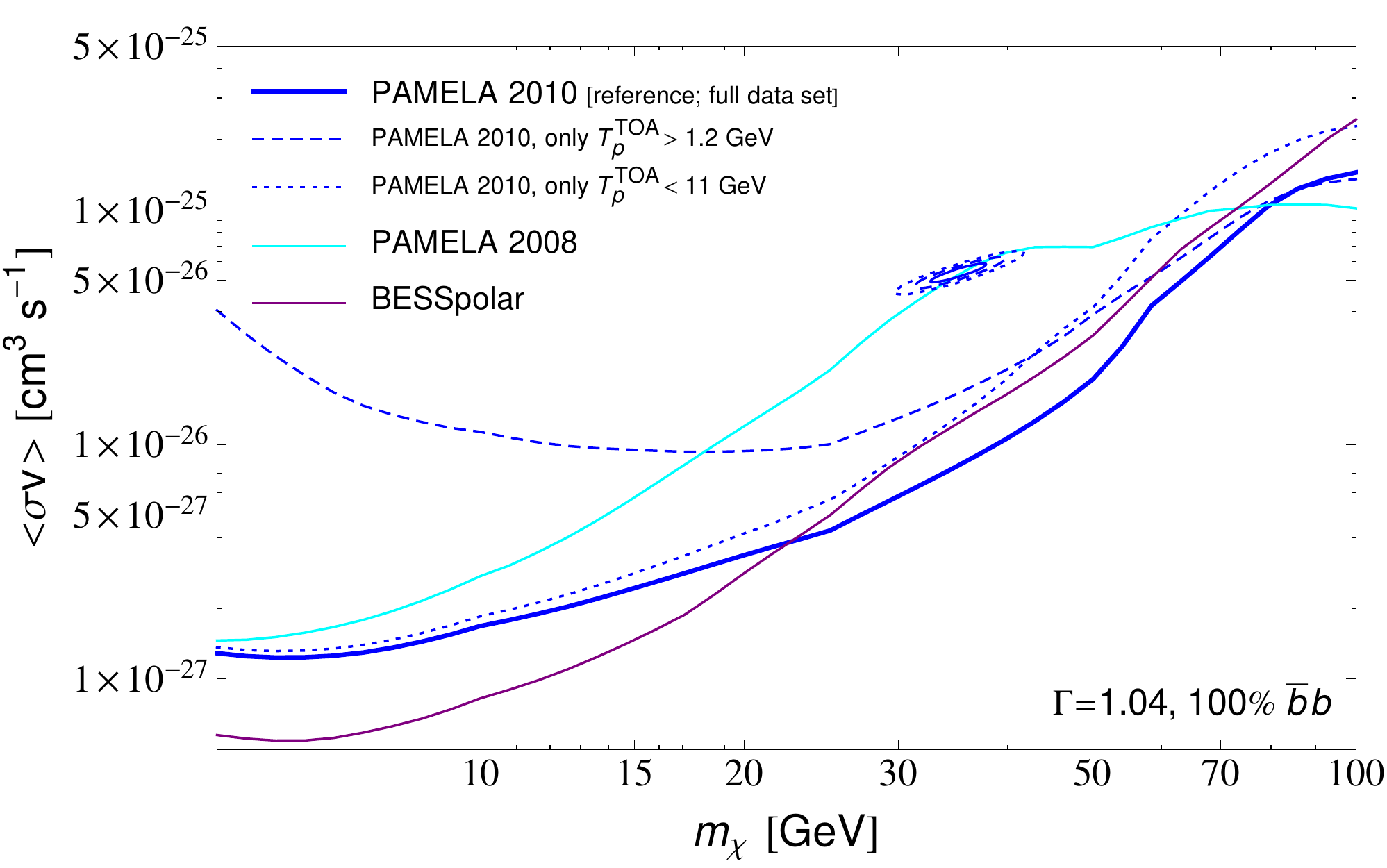}
    \end{center}
    \caption{Reference $\bar{p}$ limits and dependence on energy cuts and/or different 
    data sets (assuming an NFW-like profile, with $\Gamma=1.04$, and annihilation into 
    $\bar b b$). For comparison, we also show the claimed signal region for a DM 
    interpretation of the GC GeV excess in this channel  \cite{Daylan:2014rsa}.}
    \label{fig:pbar_systematics}
\end{figure}

As we have stressed earlier, one of the main reasons for why we could improve 
previous antiproton limits is the statistical analysis we have adopted. In particular, we 
took into account that nuclear uncertainties are not uncorrelated in energy (as was 
done \textit{e.g.}~in Ref.~\cite{Fornengo:2013xda}), while still allowing in our fits 
for the whole range of both nuclear and propagation uncertainties typically accounted 
for in the literature. Even though one might in principle still worry that these uncertainties 
may in reality be larger, in particular when considering more complicated propagation 
models, it is worthwhile to emphasize again 
that our simple 3-parameter background model provides an extremely good fit to the 
data. Despite a certain degeneracy in these three parameters, in fact, the data 
constrain the shape of the background flux extremely tightly -- which is the reason 
why we can derive strong limits in particular for light DM contributions, which feature a 
somewhat different spectral shape of the antiproton flux at low energies (for a given 
value of the Fisk potential).

As demonstrated in Fig.~\ref{fig:pbar_systematics}, it is indeed almost the full data range 
that is responsible for setting the limits, not only the lowest data bins. Even though the 
DM contribution is negligible at higher energies, \textit{e.g.}, all data points below about 
20 GeV contribute significantly to fixing {\it all} parameters 
$(\alpha_\mathrm{prop},\alpha_\mathrm{nuc},\phi_F)$. As a consequence, the DM 
contribution at these energies cannot easily be compensated by a change in those 
parameters. In fact, even neglecting data points below $T_{\bar p}\sim 1$\,GeV in the 
analysis allows to set stringent limits due to the very small error bars in the data at 
intermediate energies (unless, obviously, one considers very low DM masses).

In the same figure, we show for comparison also limits obtained with the older PAMELA 
data \cite{Adriani:2012paa} and the data taken by the BESS-Polar experiment 
\cite{Abe:2008sh}. Interestingly, the updated PAMELA analysis allows to place 
significantly stronger limits on a DM contribution exactly in the range of masses relevant 
for the GC GeV excess (in particular for the $\bar b b$ channel, for lighter quarks the 
difference is less pronounced). Notably, this is not mainly due to the longer observation 
time but due to an optimized 'spillover' analysis  \cite{Adriani:2012paa} that results in a 
somewhat different shape of the best-fit background model: in contrast to the old data, 
the new PAMELA data \cite{Adriani:2012paa} favour the maximal possible flux 
contribution that can be attributed to nuclear and propagation uncertainties 
(i.e.~$\alpha_\mathrm{prop}=\alpha_\mathrm{nuc}=1$), which is compensated by a 
significantly larger best-fit value for the Fisk potential $\phi_F$ (769\,MV rather than 
496\,MV). Despite smaller error bars, this improves the total $\chi^2$ of the best-fit 
model by more than 2. 
In view of these relatively large differences, it is comforting to see that an analysis of the 
BESS-Polar data results in limits that are comparable to the ones obtained with the 
newer PAMELA data.\footnote{
BESS-Polar only measured antiproton energies up to 4.7\,GeV, which limits the 
possibility to constrain large DM masses if the Fisk potential is left as a completely 
unconstrained parameter in the analysis. We thus imposed the very 
conservative \cite{usoskin} 
restriction of $\phi_F<1.5$\,GV in our fits, which starts to affect limits from BESS-Polar 
for $m_\chi\gtrsim 30$\,GeV, but has no effect on any of the other limits shown in 
Fig.~\ref{fig:pbar_systematics} or elsewhere.
}

\begin{figure}[t]
   \begin{center}
       \includegraphics[width=\linewidth]{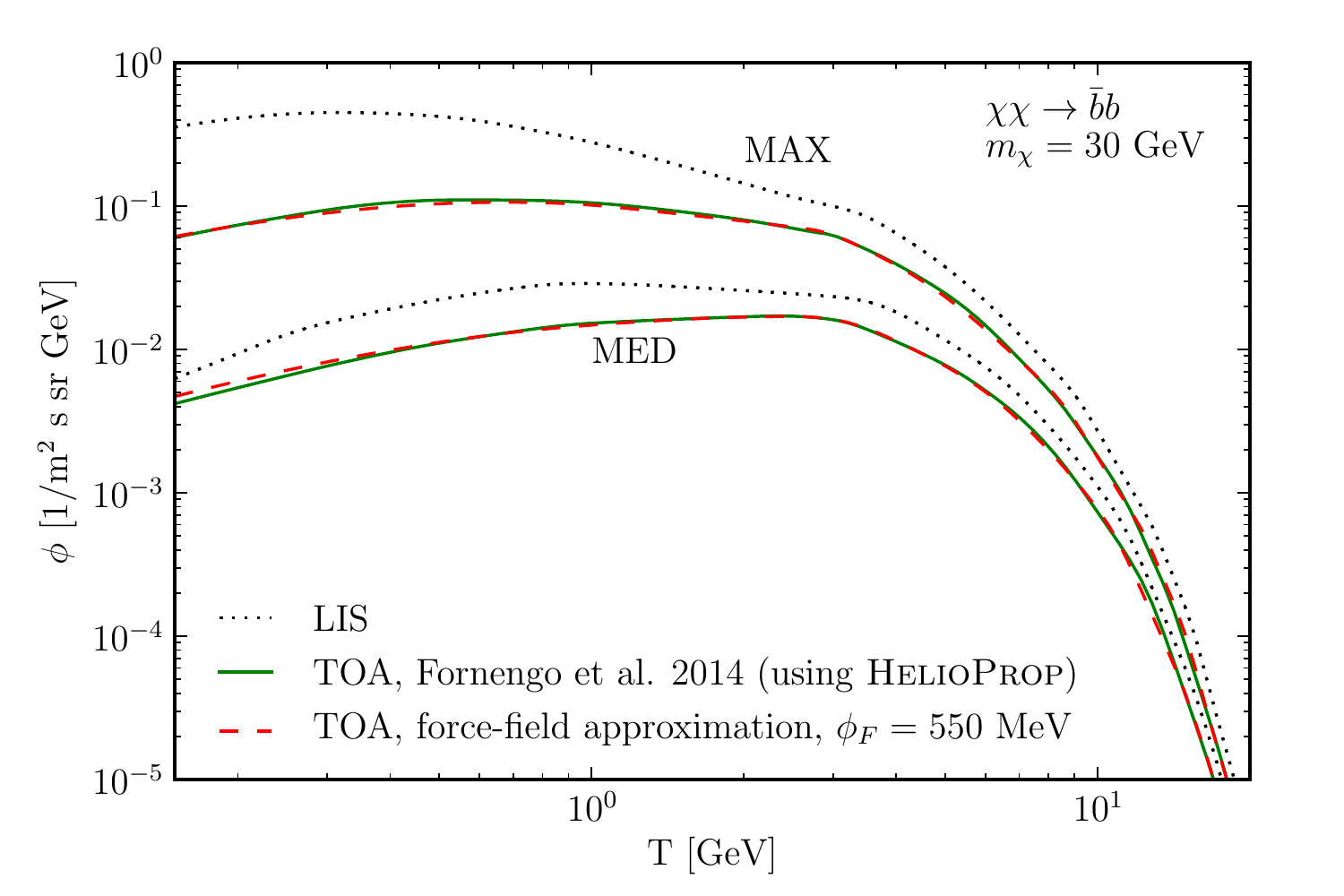}
   \end{center}
   \caption{Comparison of the force-field approximation to solar modulation
       with numerical results using \textsc{HelioProp}~\cite{Maccione:2012cu}.
       The dotted black line shows the local interstellar flux as function of
       the antiproton kinetic energy; the dashed and solid lines show the
       top-of-atmosphere flux obtained from the force-field
       approximation and the numerical results from
       Ref.~\cite{Fornengo:2013xda}.  We adopt here as two exemplary LIS fluxes the
       antiproton signals from $\chi\chi\to\bar{b}b$ with thermal cross-section,
       $m_\chi=30\rm\ GeV$, and using the MED and MAX cosmic-ray propagation
       scenarios~\cite{Donato:2003xg}, respectively.}
   \label{fig:forcefield}
\end{figure}

Another potential systematic limitation of our analysis is our treatment of solar
modulation.
The full 4D propagation equations, including the diffusion and drift motion
along the large scale gradients of the spiralling solar magnetic field, the
heliospheric current sheet, and the radially expanding solar wind, were
recently implemented in the numerical code
\textsc{HelioProp}~\cite{Maccione:2012cu}.  Ref.~\cite{Fornengo:2013xda}
employed this code to analyse in detail the effect of solar modulation on
cosmic-ray antiprotons from DM annihilation.
On the other hand, a well-known analytic solution to the effect of the heliosphere on the 
flux of cosmic rays -- the \emph{force-field 
approximation}~\cite{Gleeson:1968zza,1987AA184119P} -- is obtained under the 
simplifying assumption of spherical symmetry and constant diffusion.  It relates
the LIS to the TOA flux with a single modulation parameter, the Fisk potential
$\phi_F$, via (see Refs.~\cite{2004JGRA..109.1101C, Potgieter:2013pdj} for a
derivation and discussion)
\begin{equation}
   \frac{\Phi_{TOA}(T_{TOA})}{\Phi_{LIS}(T_{LIS})}  = \frac{p^2_{TOA}}{p^2_{LIS}}= \frac{T_{TOA}(T_{TOA} + 2m_p)}{T_{LIS}(T_{LIS} + 2m_p)}\;,
   \label{eqn:FF}
\end{equation}
with kinetic energies related by $T_{TOA} = T_{LIS} - \phi_F$.

In Fig.~\ref{fig:forcefield}, we compare the numerical results from
Ref.~\cite{Fornengo:2013xda} that were obtained with \textsc{HelioProp} for two
specific sets of propagation parameters with the results that we find by simply
applying the force-field approximation, Eq.~\eqref{eqn:FF}, like we did in our analysis.  
As benchmark
scenarios we adopt one of the channels that well reproduce the gamma-ray GeV
excess, $\chi\chi\to\bar{b}b$ with $m_\chi=30\rm \ GeV$, assuming a thermal
annihilation cross-section, and the MED and MAX scenarios \cite{Donato:2003xg}
for the antiproton
propagation in the Galaxy.  We find that, when adopting a Fisk potential of
$\phi_F = 500\rm\ MeV$, the force-field approximation reproduces the numerical
results remarkably well.  Though a detailed comparison between the force-field
approximation and a large set of heliospheric propagation parameters is still
lacking, we conclude that the force-field approximation is very well suited to
study the impact of solar modulation on DM searches with antiprotons for kinetic 
energies $T_{\bar p}\gtrsim 0.1$\,GeV.  In
particular, uncertainties related to heliospheric propagation appear negligible
w.r.t.~uncertainties coming from cosmic-ray propagation in the Galaxy.

Finally, let us remark that in the above analysis, we followed the common 
approach of adding statistical and systematic error bars in quadrature 
to obtain an estimate for the overall  flux uncertainty.  This step neglects 
possible bin-to-bin correlations between the systematic errors, which in the 
case of the PAMELA data may be overly optimistic \cite{privateComm}.  A 
correct treatment of systematic
uncertainties would require knowledge about the covariance matrix of the
systematic errors, which is however not available.  The impact on our results
would then, in principle,  critically depend on the spectral similarity between the 
principal components of the covariance matrix and the shape of signal and 
background fluxes.

While we leave a more detailed study to future work, let us here briefly illustrate 
the possible impact of correlated systematical errors on our results for a few 
simple toy scenarios.  Firstly, we allow the \emph{normalization} of the measured
flux to vary by $\pm5\%$, corresponding to the typical error stated for the PAMELA 
2010 data. We implement
this in the fit as a constrained rescaling factor, $\alpha_s \in [0.95, 1.05]$,
for the measured fluxes.  Secondly, we allow the \emph{spectral index} of the
measured flux to vary.  This is implemented as an energy-dependent prefactor
$\alpha_t(E) = 1 + \kappa\log(E/\sqrt{E_\text{mn}E_\text{min}})$, where
$|\kappa|\leq 0.05/ \log(E_\text{max}/E_\text{min})$.  The logic is here that
the \emph{maximal} deviation that we obtain at the end-points of the measured
spectrum deviates by up to $5\%$ from the nominal value. Accounting for correlated
systematics in this way, we find
that our limits would be weakened by barely more than a factor of 2, thus not affecting
our conclusions for the reference propagation model.

\subsection{Positron limits}

Systematic uncertainties that play a role when constraining DM annihilation
into leptonic final states with the positron flux measurements of AMS-02 were
discussed in detail in Ref.~\cite{Bergstrom:2013jra}.  Here, we briefly
summarize the main aspects.

The propagation of cosmic-ray electrons (and positrons) is dominated by energy
losses rather than by diffusion, namely by efficient synchrotron radiation on
the Galactic magnetic field and by inverse Compton scattering on starlight,
thermal dust radiation and the CMB.  As a consequence, positrons are a much 
more local probe of DM annihilation than the other messengers that we have 
considered here. In case of the channel that is most
constrained by AMS-02 data, annihilation into electron-positron pairs, the
resulting spectrum after propagation has a sharp step-like cutoff at energies
$E_{e^\pm}=m_\chi$, with a long tail towards lower energies.  The electron flux
at this energy, and hence the height of the step, depends almost exclusively on
the local energy-loss rate, and hence on the local radiation field, which can
be estimated to within $50\%$.  The details of the propagation scenario (height
of diffusive halo etc) play on the other hand only a marginal role for the step-like
signal.

Solar modulation will affect the TOA flux of cosmic-ray electrons
at energies below 5--10 GeV.  For DM masses around $m_\chi\approx 10\rm\ GeV$,
this is not relevant for $e^+ e^-$  and $\mu^+\mu^-$ final states, where the
relevant peak of the signal is close to the DM mass, but it can slightly affect limits
in case of the broader $\tau^+\tau^-$ spectrum.

Lastly, the discreteness of astrophysical sources that might cause the rise in
the positron fraction can potentially lead to small variations in the measured
energy spectrum.  This is currently not observed, but can lead to a variation
of the limits on $e^+e^-$ final states by up to a factor of about three.

\subsection{Radio limits}
\label{sec:radio_discussion}

Our limits in the radio band are essentially based on three main assumptions:
{\it (i)} diffusion, advection effects and energy losses other than those due
to synchrotron emission can be neglected, {\it (ii)} 
energy equipartition in
the accretion zone, i.e.~at galactocentric distances smaller than
$\sim\SI{0.04}{pc}$, and magnetic flux conservation outside, {\it (iii)} the
monochromatic approximation for synchrotron emission. As already discussed,
however, a strong magnetic field as adopted in our model is observationally well
supported~\cite{Eatough:2013nva}, and
our limits are rather insensitive to
the monochromatic approximation; we will hence mostly 
concentrate on a discussion of {\it (i)}.
\medskip

In deriving Eq.~(\ref{eq:Blimit}) as a condition for {\it (i)} to be satisfied,
we made use of the fact that for sufficiently large magnetic fields and
turbulent conditions the diffusion coefficient approaches the critical ({\it
i.e.}~minimum possible) value of $D\sim D_\text{Bohm} = r_g c/3$. 
Deviations from the assumption of Bohm diffusion will lead to less frequent
scattering and more efficient transport, $D=\kappa D_\text{Bohm}$, where we
introduced a free parameter $\kappa\geq1$ to account for this possibility.
Importantly, even allowing $\kappa\gg1$
the condition for the neglection of diffusion in Eq.~\eqref{eq:Blimit} scales only
with a factor of $\kappa^{1/3}$ on the right-hand side (at least for constant $B$).
This has profound consequences for the robustness of our radio limits:  The 
bulk of the radio signal of interest comes from a relatively small range of
distances from the Galactic center, around $r\sim 0.1$\,pc, depending on the
shape of the DM profile (see Fig.~\ref{fig:Bfield}).  Observations give a lower
limit on the magnetic field that is around $8\rm\ mG$~\cite{Eatough:2013nva} at
these scales.  We thus find that at $r\simeq 0.1$\,pc diffusion can
be neglected even if the diffusion constant is $\sim 10^7$ times larger
than the Bohm value.
\medskip

Since the requirement of \textit{in situ} energy loss is very well satisfied,
our limits depend much more on the DM profile than on the magnetic field
strength.  This is visualized in Fig.~\ref{fig:radiotautau}, but also directly 
apparent from Eq.~(\ref{eq:synflux}) where $\rho_\chi$
enters squared and gives a particularly large contribution to the volume
integration for singular profiles. The product $EN_e(E)$, on the other hand,
determines the dependence on $B$ (implying in fact slightly stronger limits for
weaker magnetic fields). In the limit of $E\ll m_\chi$, corresponding to large
magnetic field values, $N_e(E)$ approaches a constant and the synchrotron flux
scales as $E\propto B^{-1/2}$. In general, $N_e(E)$ is a monotonically
decreasing function of $E$, which implies that the actual $B$-dependence of
Eq.~(\ref{eq:synflux}) will in practice be smaller. 
An intuitive way to
understand this relatively weak dependence of the flux density for a given
frequency is that, in the limit of synchrotron losses happening {\it in situ}
as discussed above, the value of $B$ does not affect anymore the total power
radiated away but merely the frequency at which this happens. 
Note also that for larger magnetic fields $B$, we do not need to follow the
super-conservative approach of only integrating up to 1\ pc (see
Fig.~\ref{fig:Bfield}).  This implies that -- in case of cored profiles --
limits will  realistically weaken even less with increased magnetic
field values.

\begin{figure}[t]
   \begin{center}
       \includegraphics[width=\linewidth]{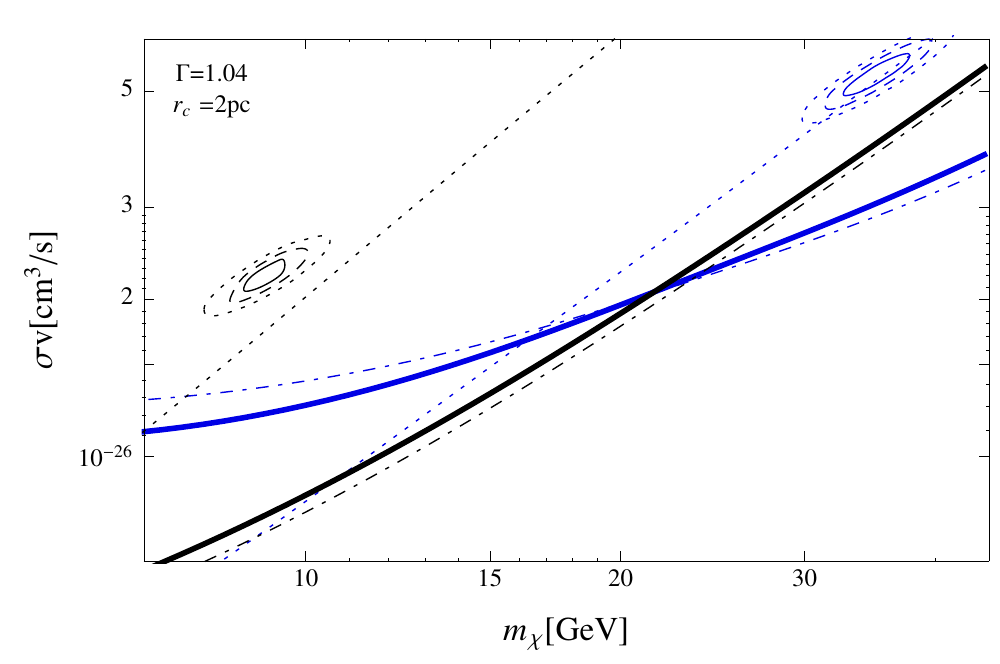}
   \end{center}
   \caption{Same as Fig.~\ref{fig:radiotautau}, but for constant magnetic
       fields of strength $B=\SI{50}{\micro G}$ (dot-dashed lines), 
       $B=\SI{8}{m G}$ (dotted lines), and our baseline
        magnetic field model (solide lines). The constant values
        correspond to the recently obtained lower limits at a galactocentric 
        distance of $\sim0.1$\,pc \cite{Eatough:2013nva}.}
   \label{fig:radio_discussion}
\end{figure}

To illustrate the scaling of the limits with $B$, we present in
Fig.~\ref{fig:radio_discussion} the cases
where the magnetic field takes the constant values $B=50\ \mu\rm
G$ and $B=8\rm\ mG$, as motivated by the recently obtained lower limits
\cite{Eatough:2013nva}  discussed above, and compare the resulting 
limits with those shown for our
baseline magnetic field model in Fig.~\ref{fig:radioallchann}. 
Interestingly, the case of a constant magnetic field of $B=50\ \mu\rm
G$ leads indeed to very similar limits. At DM masses $m_\chi\gg50$\,GeV,
limits {\it weaken} as expected with a factor of 
$(8\,\mathrm{mG}/50\,\mathrm{\mu G})^{1/2}\sim13$ when instead adopting 
the much larger constant field of $B=8\rm\ mG$. At low masses, they 
may however even {\it strengthen} due to the $N_e(E)$ dependence 
discussed above.
In both cases, we see that DM annihilation into 
80\% $\tau^+\tau^-$ + 20\% $\bar bb$ leads to limits that
exclude the interpretation of the excess in terms of such annihilations,
whereas a large magnetic field of $B=8\rm\ mG$ would require a slightly smaller
core size in the profile, $r_c\lesssim2$\,pc, to fully exclude also the 
possibility of  $\bar{b}b$ final states. 

Let us conclude this Section by mentioning a very recent 
analysis~\cite{Cholis:2014fja} which revisited radio constraints on DM 
annihilation in the GC. In particular, this analysis challenges the standard 
assumption of {\it in situ} energy losses of the emitted electrons due
to synchrotron radiation when taking into account inverse Compton 
scattering (ICS), or strong convective winds in the inner Galaxy. For the 
magnetic field model that we have adopted, e.g., the limits are claimed to 
weaken by more than two orders of magnitude when including ICS off the 
dense interstellar radiation field close to the GC (but note that the actual 
$B$-field is likely stronger, as discussed above, which would make ICS 
less important in proportion). Together with the assumption of a very strong 
convective wind that blows electrons away from the galactic  disk with a 
velocity of $v_c\sim1000$\,km/s, this would result in a weakening of our 
limits by less than 4 orders of magnitude in total. For a steep profile with 
$\Gamma=1.26$, however, even this extreme case is not sufficient to make 
the GC excess compatible with radio observations unless one introduces 
an artificial cutoff at $r_c\gtrsim0.1$\,pc.

\section{Summary}\label{sec:summary}

\begin{figure*}[t]
   \begin{center}
       \includegraphics[width=.494\linewidth]{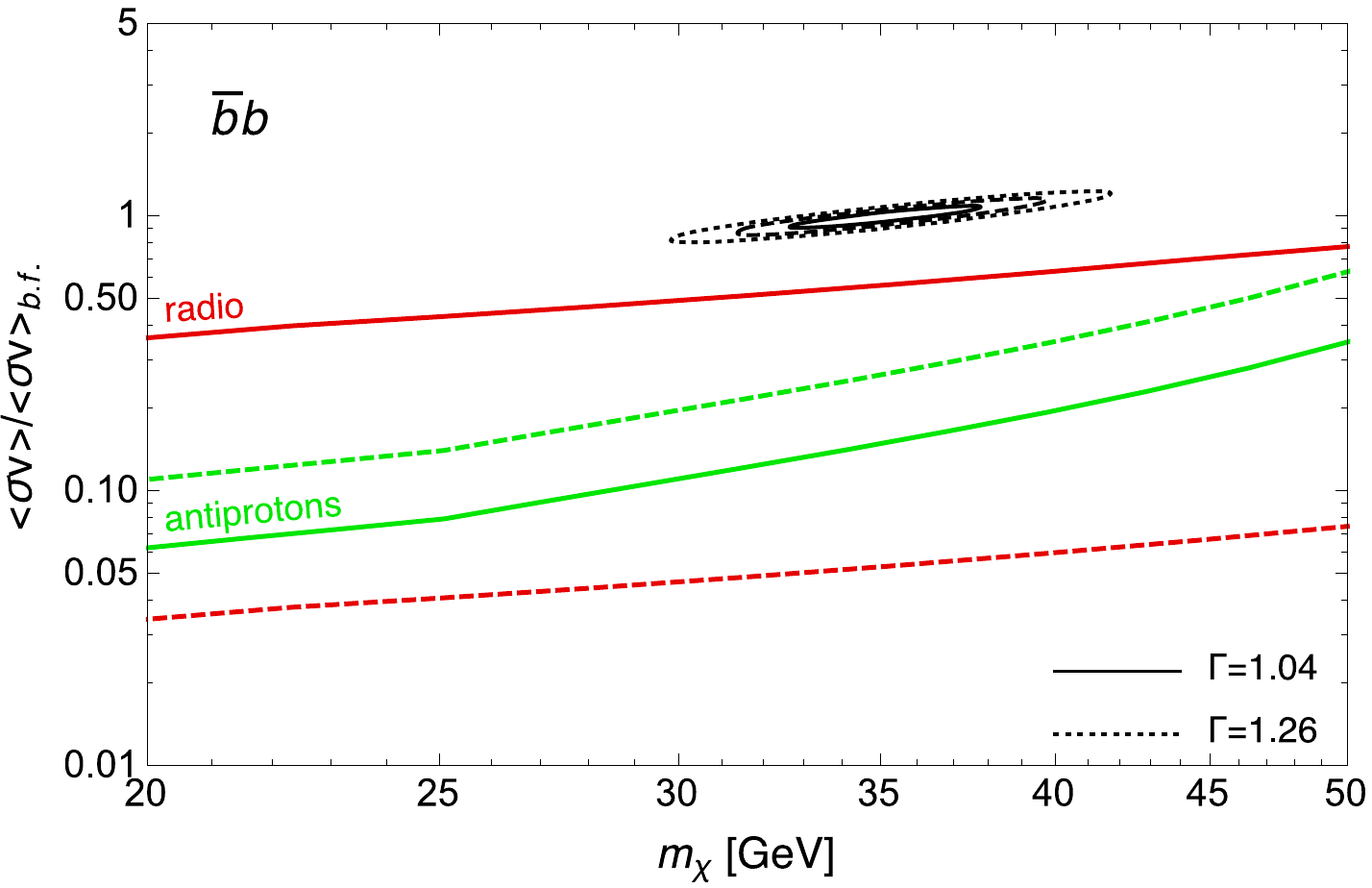}
       \includegraphics[width=.494\linewidth]{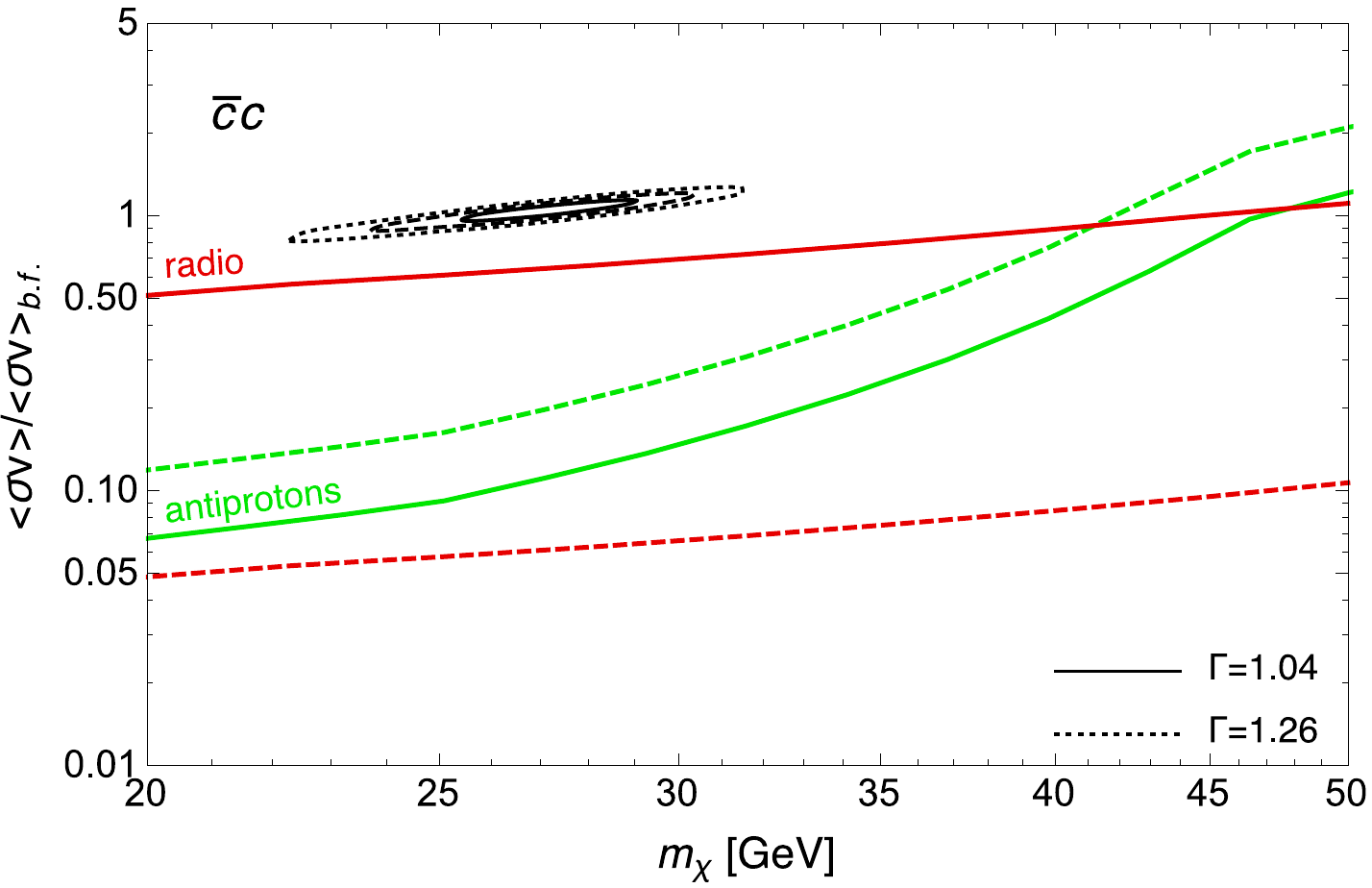}\\
       \includegraphics[width=.494\linewidth]{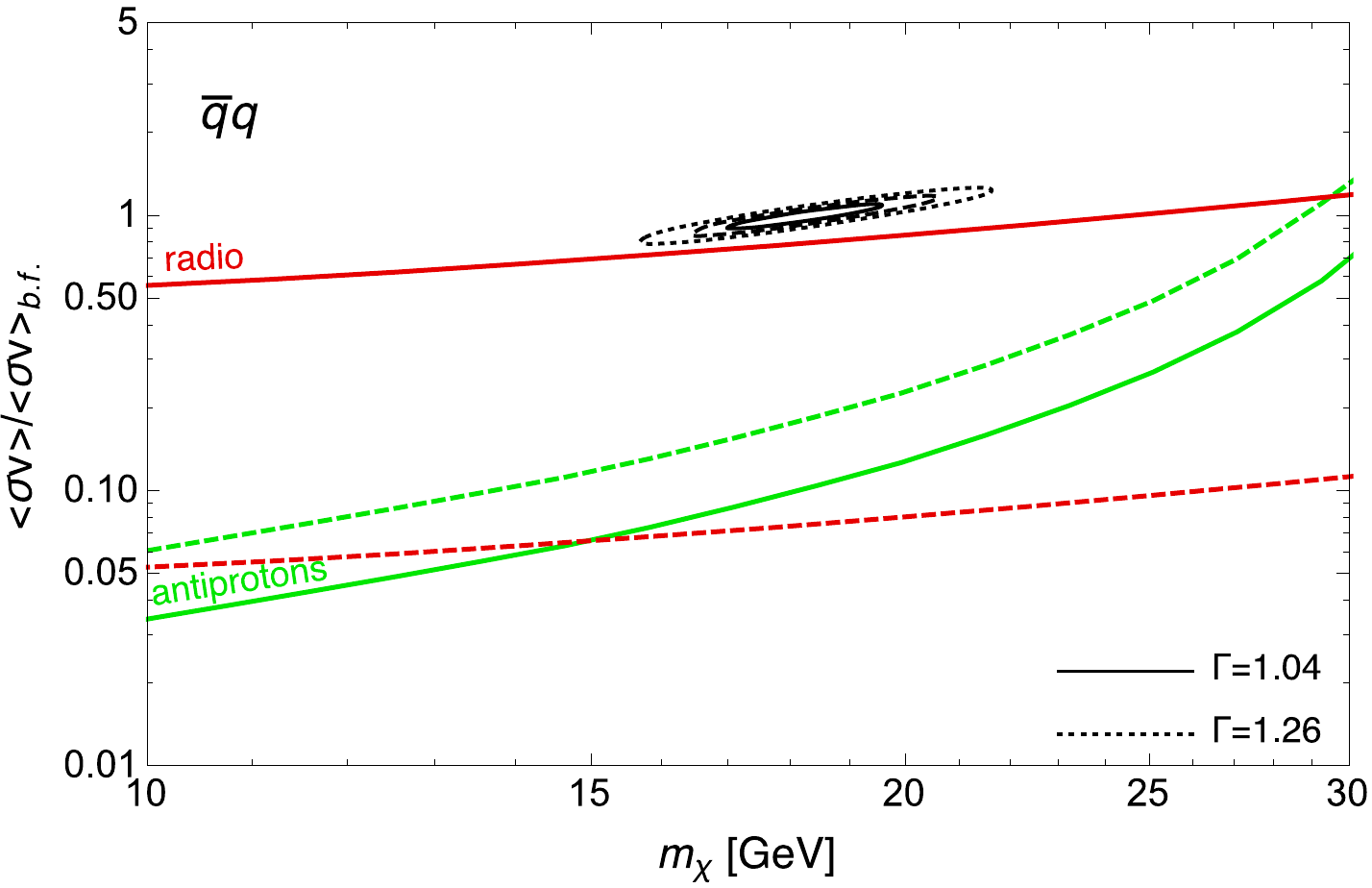}
       \includegraphics[width=.494\linewidth]{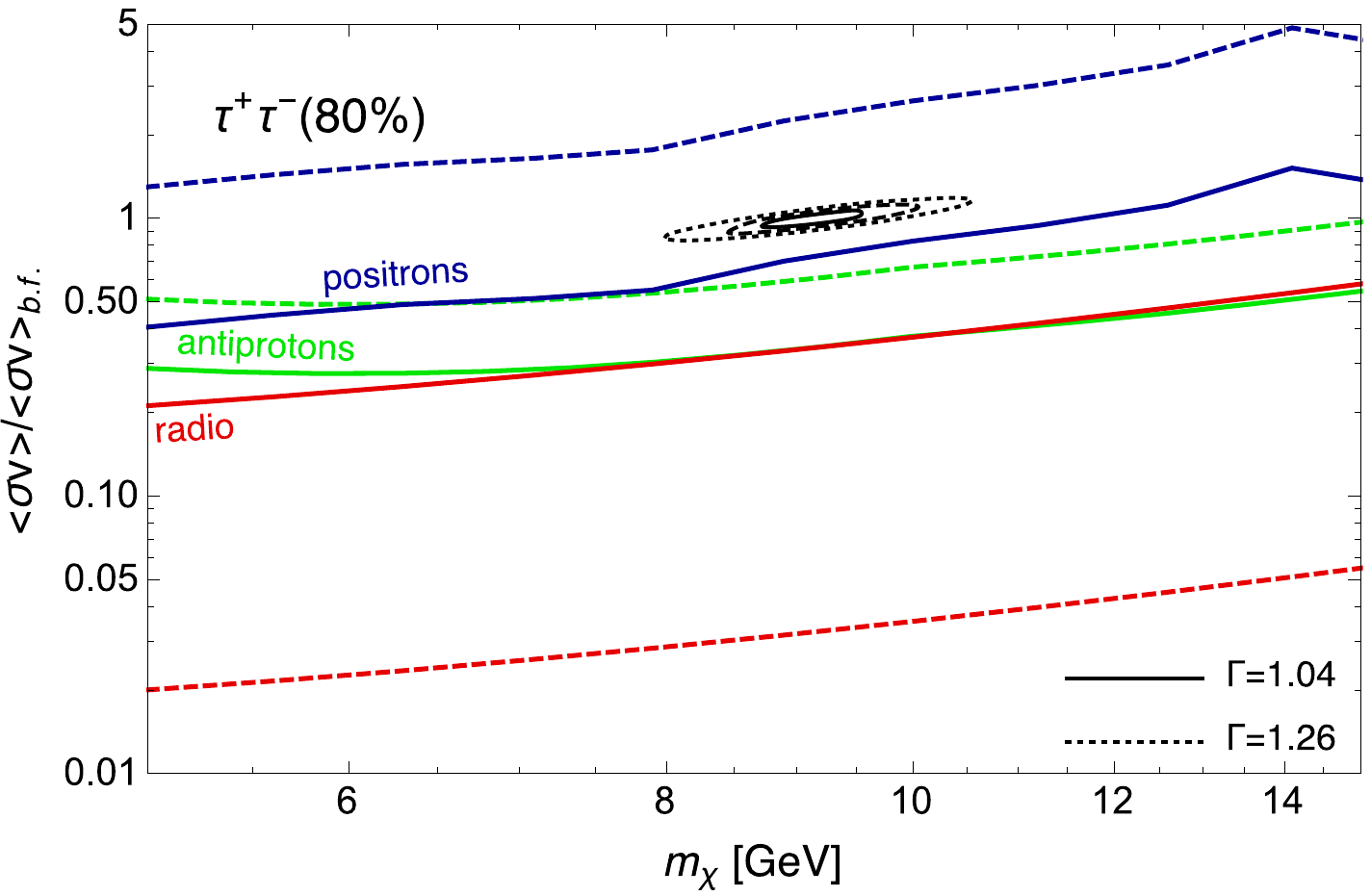}
   \end{center}
   \caption{\emph{Summary of our limits} for various annihilation channels,
   expressed in terms of the {\it ratio} of the corresponding limit on
   $\langle\sigma v\rangle$ and the best-fit DM signal interpretation. For
   convenience, we also show the corresponding 1$\sigma$, 2$\sigma$ and 3$\sigma$
   regions describing the signal from the inner Galaxy analysis
   \cite{Daylan:2014rsa} (see also Fig.~\ref{fig:cross-sections}). Dashed lines
   show limits for a generalized NFW profile with slope $\Gamma=1.26$ and solid
   lines show the corresponding limits for $\Gamma=1.04$ (note that the limits
   presented this way are independent of any overall normalization of the DM
   profile, e.g. in terms of the local DM density  $\rho_\odot=0.3\rm\ GeV\
   cm^{-3}$). In all panels, antiproton limits are displayed  in green, positron
   limits in blue and radio limits in red. Radio constraints assume that the
   DM profile flattens out at radii below 2 pc from the GC. See text for
   further details.}
   \label{fig:summary} 
\end{figure*}

In Fig.~\ref{fig:summary} we present a summary of the constraints on DM
annihilation that we derived from antiproton, positron and radio observations,
and confront them with representative benchmark scenarios that can explain the
GC GeV excess observed with Fermi LAT.  Since the annihilation cross-section
that best reproduces the gamma-ray signal depends on the assumption on the DM
profile, in particular the inner slope $\Gamma$ and the local DM density, in
Fig.~\ref{fig:summary} we normalize the annihilation cross-section to the
values that best reproduce the GeV excess
(\textit{cp.}~Tab.~\ref{tab:cross-sections}).  In this way, we collapse the
best-fit regions into one single region, which is per construction centered
onto one.  The limits change then according to the slope $\Gamma$ (and the
scale radius $r_s$, which we will comment on below), whereas the
overall normalization of the DM profile drops out (as do symmetry factors in the 
annihilation rate related to whether or not the DM particles are their own antiparticles).

In case of DM annihilation into $\bar{b}b$ final states, as shown in the upper
left panel of Fig.~\ref{fig:summary}, limits from existing observations of
antiprotons firmly exclude the DM interpretation of the GeV excess by almost one
order of magnitude, if our benchmark KRA scenario is adopted for the
propagation of Galactic cosmic rays.  These limits weaken by about a factor of 
two if $\Gamma$ increases from
$\Gamma=1.04$ to $\Gamma=1.26$.  
Note that, as shown in
Fig.~\ref{fig:gamma-dependence}, variations in the scale radius $r_s$ introduce
another uncertainty in the {\it local} annihilation rate that can be as large as a factor 
of two, which in case of very cuspy profiles (like $\Gamma=1.26$) will typically tend to
\emph{increase} the annihilation signal when mass constraints on the Milky Way
are taken into account. However, we checked explicitly that the antiproton
limits -- which probe a relatively large annihilation volume -- are only strengthened 
by 24\,\% (weakened by 39\,\%) when changing our references value of 
$r_s$ from 20\,kpc to 35\,kpc (10\,kpc).
As shown in Fig.~\ref{fig:pbar_limits_prop},
in case of the very conservative MIN' propagation scenario the limits can
weaken by an additional factor of 7--8.  Only in this extreme propagation setting, the 
GeV excess would still be marginally consistent with current antiproton observations
-- implying that AMS-02 will either observe a corresponding excess in antiprotons or 
rule out even this remaining possibility.  

We also show limits obtained from 408\,MHz Jodrell Bank radio observations of the
inner 4'' centered on the GC.  As discussed above, these limits are rather robust
concerning assumptions on the magnetic field, but they critically depend on the
adopted DM profile.  Most importantly, in case of the gamma-ray excess, the DM
profile is fixed to a large extent, allowing to make much more precise signal
predictions.  For a profile that remains cuspy all the way down to $\lesssim 0.1\rm\
pc$, all channels that could explain the GeV excess are ruled out by several orders 
of magnitude~(\textit{cp.}~Fig.~\ref{fig:radiotautau}) -- though these limits become much
less severe when allowing for strong convective winds or ICS to dominate over 
synchrotron emission \cite{Cholis:2014fja}. 
In Fig.~\ref{fig:summary}, we show
limits that are obtained for an \textit{ad hoc} flattening of the generalized NFW profile
at radii below 2 pc.  Even with such an artificial core, in case of the cuspy
$\Gamma=1.26$ the limits still exclude the DM interpretation of the GeV excess
by more than one order of magnitude.
Note that GC radio observations only probe the innermost region of the DM halo.
The resulting limits thus do not depend on the scale radius $r_s$, 
see Fig.~\ref{fig:gamma-dependence}. For the same reason, steeper profiles 
imply much {\it stronger} radio constraints  in the context of the GeV excess,
while they imply slightly {\it weaker} constraints for antiprotons. 
In case of DM annihilation into $\bar{c}c$ or light quarks $\bar{q}q$ (upper
right and lower left panel of Fig.~\ref{fig:summary}), the
limits obtained from radio and antiproton observations are quantitatively very similar to 
the case of $\bar{b}b$.  

In the lower right panel of Fig.~\ref{fig:summary}, we finally show limits on
the mixed final state $\tau^+\tau^-$ (80\%) plus $\bar{b}b$ (20\%), which was
motivated in Ref.~\cite{Daylan:2014rsa} by a spectral fit to the excess seen in
the inner Galaxy (see also the discussion in Section \ref{sec:gevDM}).  In this 
case, antiproton constraints only lead to an exclusion by a factor of roughly 2 in case 
of the KRA propagation scenario (and no exclusion if the MIN' model is 
adopted).  However, radio limits firmly exclude this channel for a DM profile 
that remains cuspy down to at least 2 pc.  
Additional robust constraints derive from the shape analysis \cite{Bergstrom:2013jra} 
of the flux of cosmic-ray positrons as measured by AMS-02, which again marginally 
excludes the interpretation of the gamma-ray GeV excess in terms of DM annihilation 
into $\tau^+\tau^-$ final states for a cuspy profile with $\Gamma=1.26$ (as preferred
by the inner galaxy analysis of Ref.~\cite{Daylan:2014rsa}). 
 As discussed above (see Figs.~\ref{fig:triangle}
and~\ref{fig:triangle2}), these limits are \emph{much} stronger in case of
other leptonic final states, and exclude branching ratios into $e^+e^-$ down to
about $\sim2.3\times 10^{-2}$, and down to $\sim 2.5\times 10^{-1}$ in case of
$\mu^+\mu^-$ final states. 
Note that uncertainties related to the scale radius $r_s$ of the DM profile are larger 
than for antiprotons and can in principle be as large as a factor
of two; for $\Gamma=1.26$, however, they  tend to {\it increase} the local annihilation
signal. 

We recall that a larger contribution of $\tau\tau$ final
states than 80\% results in a spectrum that no longer provides a reasonable fit
to the observed GeV excess, even when taking into account the contribution from 
bremsstrahlung and inverse Compton scattering. A smaller contribution, on the 
other hand, necessarily implies a larger $\bar bb$ (or lighter quark) contribution 
in view of the strong AMS-02 limits on light leptons as annihilation products. While 
a mixed final state with a smaller $\tau^\pm$ component could evade AMS-02 
bounds, the antiproton bounds from PAMELA would thus become even more severe.
\bigskip

\emph{In summary}, we find that the DM interpretation of the GeV excess in
terms of hadronic or leptonic final states is strongly constrained by our
analysis of existing antiproton, positron and radio data.  
A way to make these limits even more robust would be to study diffusion parameters, 
in particular the truly minimal 
diffusion zone height, in light of existing radio, gamma-ray and cosmic-ray 
observations in a \emph{combined} analysis. For the case of antiproton limits, the 
most important next step would then be to systematically improve on the nuclear 
uncertainties connected to secondary antiproton production. For positron limits 
on $\tau^+\tau^-$ final states, on the other hand, a dedicated study of the effects of solar 
modulation on the spectrum of DM  induced positrons below energies of 
$\sim 5\rm\GeV$ would be more warranted. Avoiding the radio limits we have presented 
requires a dedicated discussion of what mechanisms could give rise to a DM profile that 
is cuspy in the inner Galaxy, down to $\mathcal{O}(10)$\,pc scales, but has a core with 
an extension of a few pc at its center. 
Last but not least, it will certainly help to get a better handle on the systematics 
connected to the determination of the DM profile slope $\Gamma$, as well as the scale 
radius $r_s$, from gamma-ray and dynamical observations.

While all  those possible directions of further investigation clearly lie beyond the scope 
of the present study, they certainly indicate that there is room to make indirect DM 
searches even more competitive -- especially for light DM models, but independently of
the concrete application of the GeV gamma-ray excess currently claimed at the GC
and inner Galaxy.

\section{Conclusions}\label{sec:concl}

In this article, we have revisited current bounds from indirect searches for DM and 
identified the messengers and targets that are most relevant for light DM, i.e.~for 
masses roughly below 100\,GeV:
\begin{itemize}
\item
 For DM annihilation into light charged leptons, {\it positrons} provide the most stringent 
 and robust bounds \cite{Bergstrom:2013jra, Ibarra:2013zia}. This constrains the 
 'thermal' annihilation rate of 
 $\langle\sigma v\rangle=3\cdot10^{-26}$cm$^3$s$^{-1}$ for masses of roughly 
 $m_\chi\lesssim 200$\,GeV ($m_\chi\lesssim 100$\,GeV) assuming dominant 
 annihilation to $e^+e^-$ ($\mu^+\mu^-$) final states. 
 \item
 As has been pointed out earlier, {\it antiprotons} provide a very efficient means of 
 constraining light DM annihilating into quarks \cite{Bottino:1998tw,Bottino:2005xy,
 Lavalle:2010yw,Evoli:2011id}. Here, we have derived new  
 bounds that constrain the thermal annihilation rate for $m_\chi\lesssim 55$\,GeV  in the 
 case of $\bar bb$ final states, and for $m_\chi\lesssim 35$\,GeV if annihilation into light 
 quarks dominates. 
\item
Null-searches for 408\,MHz {\it radio signals} from a 4'' region around the GC, finally, 
provide extremely stringent constraints \cite{Gondolo:2000pn,Bertone:2001jv,Aloisio:2004hy,Regis:2008ij,
Bertone:2008xr,Bringmann:2009ca,Mambrini:2012ue,Laha:2012fg,Asano:2012zv}.  We revisited those bounds and 
discussed that the dependence on assumptions about the magnetic field is typically much smaller
than  that related to the DM density at pc scales away from the GC. For a cuspy 
profile (like NFW) that extends down to at least  0.1\,pc, thermal cross sections are 
excluded for DM masses below roughly 120 GeV (for $\tau^+\tau^-$) or even 400 GeV
(for $\bar b b$). 
\end{itemize}

As an application, we have discussed at length that these findings are particularly 
relevant for a possible DM interpretation of the much-debated excess in GeV gamma 
rays from the GC 
and inner Galaxy \cite{Daylan:2014rsa,Goodenough:2009gk, Hooper:2010mq, 
Hooper:2011ti, Abazajian:2012pn,
Macias:2013vya, Abazajian:2014fta}. In fact, such an interpretation requires a well-defined 
and highly constrained region both in the $\langle \sigma v\rangle$ vs.~$m_\chi$ plane 
and in the range of possible DM density profiles. This implies that constraints from other 
indirect detection methods can directly be applied and are subject to much less severe 
uncertainties than in the absence of a signal indication. For example, these constraints 
become independent of the overall normalization of the annihilation rate and thus, e.g., 
of the DM production mechanism.
In fact, probing the same
annihilation mechanism, such constraints are much more model-independent than 
constraints derived from collider or direct searches \cite{Alves:2014yha,Berlin:2014tja,
Izaguirre:2014vva,Kong:2014haa,Han:2014nba}.

For reasonable benchmark scenarios for cosmic-ray propagation and a DM density 
profile consistent with the observed excess, we basically find that all annihilation 
channels that were considered in Ref.~\cite{Daylan:2014rsa} are ruled out (the same
holds for purely leptonic annihilation channels, which were suggested in 
Ref.~\cite{Lacroix:2014eea}).  The tension can be
somewhat alleviated by (a) assuming a borderline propagation scenario with
minimal diffusion zone height (our MIN') \emph{and} (b) assuming that the DM
profile cuts off at radii of at least $\sim 5\rm\ pc$ from the Galactic
center (while keeping its $\Gamma=1.26$ slope observed at larger radii 
$r\gtrsim10$\,pc). On the model-building side, the tension could also be made somewhat 
less severe by considering cascade decays \cite{Martin:2014sxa,Boehm:2014bia,Detmold:2014qqa} rather 
than the direct decay into two SM particle final states that we have considered here. 
While a thorough investigation of this possibility lies beyond the scope of this work, we 
do not expect our conclusions to change qualitatively.

A confirmation of the DM interpretation of the GeV excess clearly requires
corroborating evidence from multiple observations.  While this could  also come 
from collider or direct probes, indirect searches arguably provide the most
model-independent way of testing the signal. Even though the excess is already in rather 
strong tension with these searches, it is conceivable that the individual uncertainties of 
the respective  limits (as discussed in Section \ref{sec:discussion} and summarized  in 
Section \ref{sec:summary})  conspire such that a signal interpretation would still be 
viable. In this case, an independent confirmation of the GeV excess as a DM 
signal -- if indeed true -- is likely to be just around the corner. 

\vspace*{1cm}

\begin{acknowledgments}
We would like to thank Lars Bergstr\"om, Celine Boehm, Mirko Boezio, Carmelo Evoli, 
Dan Hooper, Tim Linden, Meng Su, Jacco Vink, and Wei Xue for very useful discussions. 
TB acknowledges support from the German Research 
Foundation (DFG) through the Emmy Noether grant \mbox{BR 3954/1-1}. TB and CW thank 
NORDITA for generous support and an inspiring atmosphere in the context of the 
programme ``What is Dark Matter?'', where part of this work was completed. MV 
acknowledges support from the Forschungs- und Wissenschaftsstiftung Hamburg 
through the program ``Astroparticle Physics with Multiple Messengers''. This work makes 
use of SciPy \cite{scipy}, Minuit \cite{James:1975dr} and Matplotlib \cite{matplotlib}.
\end{acknowledgments}

\vspace*{1cm}

\paragraph*{Note added.} After the submission of this work, another comprehensive
analysis of antiproton constraints on the GeV excess appeared on the 
arXiv~\cite{Cirelli:2014lwa}, adopting a somewhat different background parameterization and
statistical method.

\end{document}